\documentstyle[11pt,amssymbols]{article}

\def\journal{\topmargin .3in	\oddsidemargin .5in
	\headheight 0pt	\headsep 0pt
	\textwidth 5.625in 
	\textheight 8.25in 
	\marginparwidth 1.5in
	\parindent 2em
	\parskip .5ex plus .1ex		\jot = 1.5ex}
\journal
\def\baselinestretch{1.2}
\catcode`\@=11
\def\marginnote#1{}
\newcount\hour
\newcount\minute
\newtoks\amorpm
\hour=\time\divide\hour by60
\minute=\time{\multiply\hour by60 \global\advance\minute by-\hour}
\edef\standardtime{{\ifnum\hour<12 \global\amorpm={am}%
	\else\global\amorpm={pm}\advance\hour by-12 \fi
	\ifnum\hour=0 \hour=12 \fi
	\number\hour:\ifnum\minute<10 0\fi\number\minute\the\amorpm}}
\edef\militarytime{\number\hour:\ifnum\minute<10 0\fi\number\minute}
\def\draftlabel#1{{\@bsphack\if@filesw {\let\thepage\relax
   \xdef\@gtempa{\write\@auxout{\string
      \newlabel{#1}{{\@currentlabel}{\thepage}}}}}\@gtempa
   \if@nobreak \ifvmode\nobreak\fi\fi\fi\@esphack}
	\gdef\@eqnlabel{#1}}
\def\@eqnlabel{}
\def\@vacuum{}
\def\draftmarginnote#1{\marginpar{\raggedright\scriptsize\tt#1}}
\def\draft{\oddsidemargin -.5truein
	\def\@oddfoot{\sl preliminary draft \hfil
	\rm\thepage\hfil\sl\today\quad\militarytime}
	\let\@evenfoot\@oddfoot	\overfullrule 3pt
	\let\label=\draftlabel
	\let\marginnote=\draftmarginnote
   \def\@eqnnum{(\theequation)\rlap{\kern\marginparsep\tt\@eqnlabel}%
\global\let\@eqnlabel\@vacuum}  }
\def\preprint{\twocolumn\sloppy\flushbottom\parindent 2em
	\leftmargini 2em\leftmarginv .5em\leftmarginvi .5em
	\oddsidemargin -.5in	\evensidemargin -.5in
	\columnsep .4in	\footheight 0pt
	\textwidth 10in	\topmargin  -.4in
	\headheight 12pt \topskip .4in
	\textheight 7.1in \footskip 0pt
	\def\@oddhead{\thepage\hfil\addtocounter{page}{1}\thepage}
	\let\@evenhead\@oddhead	\def\@oddfoot{}	\def\@evenfoot{} }
\def\proceedings{\pagestyle{empty}
	\oddsidemargin .26in \evensidemargin .26in
	\topmargin .27in	\textwidth 145mm
	\parindent 12mm	\textheight 225mm
	\headheight 0pt	\headsep 0pt
	\footskip 0pt	\footheight 0pt}
\def\numberbysection{\@addtoreset{equation}{section}
	\def\theequation{\thesection.\arabic{equation}}}
\def\underline#1{\relax\ifmmode\@@underline#1\else
	$\@@underline{\hbox{#1}}$\relax\fi}
\def\titlepage{\@restonecolfalse\if@twocolumn\@restonecoltrue\onecolumn
     \else \newpage \fi \thispagestyle{empty}\c@page\z@
	\def\thefootnote{\fnsymbol{footnote}} }
\def\endtitlepage{\if@restonecol\twocolumn \else \newpage \fi
	\def\thefootnote{\arabic{footnote}}
	\setcounter{footnote}{0}}  
\catcode`@=12
\relax
\def\figcap{\section*{Figure Captions\markboth
	{FIGURECAPTIONS}{FIGURECAPTIONS}}\list
	{Figure \arabic{enumi}:\hfill}{\settowidth\labelwidth{Figure 999:}
	\leftmargin\labelwidth
	\advance\leftmargin\labelsep\usecounter{enumi}}}
\let\endfigcap\endlist \relax
\def\tablecap{\section*{Table Captions\markboth
	{TABLECAPTIONS}{TABLECAPTIONS}}\list
	{Table \arabic{enumi}:\hfill}{\settowidth\labelwidth{Table 999:}
	\leftmargin\labelwidth
	\advance\leftmargin\labelsep\usecounter{enumi}}}
\let\endtablecap\endlist \relax
\def\reflist{\section*{References\markboth
	{REFLIST}{REFLIST}}\list
	{[\arabic{enumi}]\hfill}{\settowidth\labelwidth{[999]}
	\leftmargin\labelwidth
	\advance\leftmargin\labelsep\usecounter{enumi}}}
\let\endreflist\endlist \relax
\makeatletter
\newcounter{pubctr}
\def\publist{\@ifnextchar[{\@publist}{\@@publist}}
\def\@publist[#1]{\list
	{[\arabic{pubctr}]\hfill}{\settowidth\labelwidth{[999]}
	\leftmargin\labelwidth
	\advance\leftmargin\labelsep
	\@nmbrlisttrue\def\@listctr{pubctr}
	\setcounter{pubctr}{#1}\addtocounter{pubctr}{-1}}}
\def\@@publist{\list
	{[\arabic{pubctr}]\hfill}{\settowidth\labelwidth{[999]}
	\leftmargin\labelwidth
	\advance\leftmargin\labelsep
	\@nmbrlisttrue\def\@listctr{pubctr}}}
\let\endpublist\endlist \relax
\makeatother
\def\mdot{\hskip -.1cm \cdot \hskip -.1cm}
\def\slash{\not\!}
\catcode`\@=11
\def\section{\@startsection {section}{1}{0pt}{-3.5ex plus -1ex minus
 -.2ex}{2.3ex plus .2ex}{\raggedright\large\bf}}
\catcode`\@=12
\def\mbf#1{\hbox{\boldmath $#1$}}
\newskip\humongous \humongous=0pt plus 1000pt minus 1000pt
\def\caja{\mathsurround=0pt}
\def\eqalign#1{\,\vcenter{\openup1\jot \caja
	\ialign{\strut \hfil$\displaystyle{##}$&$
	\displaystyle{{}##}$\hfil\crcr#1\crcr}}\,}
\newif\ifdtup
\def\panorama{\global\dtuptrue \openup1\jot \caja
	\everycr{\noalign{\ifdtup \global\dtupfalse
	\vskip-\lineskiplimit \vskip\normallineskiplimit
	\else \penalty\interdisplaylinepenalty \fi}}}
\def\eqalignno#1{\panorama \tabskip=\humongous
	\halign to\displaywidth{\hfil$\displaystyle{##}$
	\tabskip=0pt&$\displaystyle{{}##}$\hfil
	\tabskip=\humongous&\llap{$##$}\tabskip=0pt
	\crcr#1\crcr}}
\def\oldrefledge{\hangindent3\parindent}
\def\oldreffmt#1{\rlap{[#1]} \hbox to 2\parindent{}}
\def\oldref#1{\par\noindent\oldrefledge \oldreffmt{#1}
	\ignorespaces}
\def\figledge{\hangindent=1.25in}
\def\figfmt#1{\rlap{Figure {#1}} \hbox to 1in{}}
\def\fig#1{\par\noindent\figledge \figfmt{#1}
	\ignorespaces}
\def\ie{\hbox{\it i.e.}}	\def\etc{\hbox{\it etc.}}
\def\eg{\hbox{\it e.g.}}	\def\cf{\hbox{\it cf.}}
\def\etal{\hbox{\it et al.}}
\def\dash{\hbox{---}}
\def\cok{\mathop{\rm cok}}
\def\tr{\mathop{\rm tr}}
\def\Tr{\mathop{\rm Tr}}
\def\Im{\mathop{\rm Im}}
\def\Re{\mathop{\rm Re}}
\def\bR{\mathop{\bf R}}
\def\bC{\mathop{\bf C}}
\def\lie{\hbox{\it \$}}	
\def\partder#1#2{{\partial #1\over\partial #2}}
\def\secder#1#2#3{{\partial^2 #1\over\partial #2 \partial #3}}
\def\bra#1{\left\langle #1\right|}
\def\ket#1{\left| #1\right\rangle}
\def\VEV#1{\left\langle #1\right\rangle}
\let\vev\VEV
\def\gdot#1{\rlap{$#1$}/}
\def\abs#1{\left| #1\right|}
\def\pri#1{#1^\prime}
\def\ltap{\raisebox{-.4ex}{\rlap{$\sim$}} \raisebox{.4ex}{$<$}}
\def\gtap{\raisebox{-.4ex}{\rlap{$\sim$}} \raisebox{.4ex}{$>$}}
\def\contract{\makebox[1.2em][c]{
	\mbox{\rule{.6em}{.01truein}\rule{.01truein}{.6em}}}}
\def\half{{1\over 2}}
\def\beq{\begin{equation}}
\def\eeq{\end{equation}}
\def\ul{\underline}
\def\bea{\begin{eqnarray}}
\def\lrover#1{
	\raisebox{1.3ex}{\rlap{$\leftrightarrow$}} \raisebox{ 0ex}{$#1$}}
\def\com#1#2{
	\left[#1, #2\right]}
\def\eea{\end{eqnarray}}
\def\bentarrow{\:\raisebox{1.3ex}{\rlap{$\vert$}}\!\rightarrow}
\def\longbent{\:\raisebox{3.5ex}{\rlap{$\vert$}}\raisebox{1.3ex}%
	{\rlap{$\vert$}}\!\rightarrow}
\def\onedk#1#2{
	\begin{equation}
	\begin{array}{l}
	 #1 \\
	 \bentarrow #2
	\end{array}
	\end{equation}
		}
\def\dk#1#2#3{
	\begin{equation}
	\begin{array}{r c l}
	#1 & \rightarrow & #2 \\
	 & & \bentarrow #3
	\end{array}
	\end{equation}
		}
\def\dkp#1#2#3#4{
	\begin{equation}
	\begin{array}{r c l}
	#1 & \rightarrow & #2#3 \\
	 & & \phantom{\; #2}\bentarrow #4
	\end{array}
	\end{equation}
		}
\def\bothdk#1#2#3#4#5{
	\begin{equation}
	\begin{array}{r c l}
	#1 & \rightarrow & #2#3 \\
	 & & \:\raisebox{1.3ex}{\rlap{$\vert$}}\raisebox{-0.5ex}{$\vert$}%
	\phantom{#2}\!\bentarrow #4 \\
	 & & \bentarrow #5
	\end{array}
	\end{equation}
		}
\def\ap#1#2#3{           {\it Ann. Phys. (NY) }{\bf #1}, #2 (19#3)}
\def\apj#1#2#3{          {\it Astrophys. J. }{\bf #1}, #2 (19#3)}
\def\apjl#1#2#3{         {\it Astrophys. J. Lett. }{\bf #1}, #2 (19#3)}
\def\app#1#2#3{          {\it Acta Phys. Polon. }{\bf #1}, #2 (19#3)}
\def\ar#1#2#3{     {\it Ann. Rev. Nucl. and Part. Sci. }{\bf #1}, #2 (19#3)}
\def\com#1#2#3{          {\it Comm. Math. Phys. }{\bf #1}, #2 (19#3)}
\def\comp#1#2#3{          {\it Comp. Phys. Comm. }{\bf #1}, #2 (19#3)}
\def\ib#1#2#3{           {\it ibid. }{\bf #1}, #2 (19#3)}
\def\nat#1#2#3{          {\it Nature (London) }{\bf #1}, #2 (19#3)}
\def\nc#1#2#3{           {\it Nuovo Cim.  }{\bf #1}, #2 (19#3)}
\def\np#1#2#3{           {\it Nucl. Phys. }{\bf #1}, #2 (19#3)}
\def\pl#1#2#3{           {\it Phys. Lett. }{\bf #1}, #2 (19#3)}
\def\pr#1#2#3{           {\it Phys. Rev. }{\bf #1}, #2 (19#3)}
\def\prep#1#2#3{         {\it Phys. Rep. }{\bf #1}, #2 (19#3)}
\def\prl#1#2#3{          {\it Phys. Rev. Lett. }{\bf #1}, #2 (19#3)}
\def\pro#1#2#3{          {\it Prog. Theor. Phys. }{\bf #1}, #2 (19#3)}
\def\rmp#1#2#3{          {\it Rev. Mod. Phys. }{\bf #1}, #2 (19#3)}
\def\sp#1#2#3{           {\it Sov. Phys.-Usp. }{\bf #1}, #2 (19#3)}
\def\sjn#1#2#3{           {\it Sov. J. Nucl. Phys. }{#1}, #2 (19#3)}
\def\srv#1#2#3{           {\it Surv. High Energy Phys. }{#1}, #2 (19#3)}
\def\tp{these proceedings}
\def\zp#1#2#3{           {\it Zeit. fur Physik }{\bf #1}, #2 (19#3)}
\catcode`\@=11
\def\eqnarray{\stepcounter{equation}\let\@currentlabel=\theequation
\global\@eqnswtrue
\global\@eqcnt\z@\tabskip\@centering\let\\=\@eqncr
\gdef\@@fix{}\def\eqno##1{\gdef\@@fix{##1}}%
$$\halign to \displaywidth\bgroup\@eqnsel\hskip\@centering
  $\displaystyle\tabskip\z@{##}$&\global\@eqcnt\@ne
  \hskip 2\arraycolsep \hfil${##}$\hfil
  &\global\@eqcnt\tw@ \hskip 2\arraycolsep $\displaystyle\tabskip\z@{##}$\hfil
   \tabskip\@centering&\llap{##}\tabskip\z@\cr}

\def\@@eqncr{\let\@tempa\relax
    \ifcase\@eqcnt \def\@tempa{& & &}\or \def\@tempa{& &}
      \else \def\@tempa{&}\fi
     \@tempa \if@eqnsw\@eqnnum\stepcounter{equation}\else\@@fix\gdef\@@fix{}\fi
     \global\@eqnswtrue\global\@eqcnt\z@\cr}

\catcode`\@=12
\font\tenbifull=cmmib10 
\font\tenbimed=cmmib10 scaled 800
\font\tenbismall=cmmib10 scaled 666
\def\bmit{\fam9 }
\def\boldalpha{\fam=9{\mathchar"710B } }
\def\boldbeta{\fam=9{\mathchar"710C } }
\def\boldgamma{\fam=9{\mathchar"710D } }
\def\bolddelta{\fam=9{\mathchar"710E } }
\def\boldepsilon{\fam=9{\mathchar"710F } }
\def\boldzeta{\fam=9{\mathchar"7110 } }
\def\boldeta{\fam=9{\mathchar"7111 } }
\def\boldtheta{\fam=9{\mathchar"7112 } }
\def\boldiota{\fam=9{\mathchar"7113 } }
\def\boldkappa{\fam=9{\mathchar"7114 } }
\def\boldlambda{\fam=9{\mathchar"7115 } }
\def\boldmu{\fam=9{\mathchar"7116 } }
\def\boldnu{\fam=9{\mathchar"7117 } }
\def\boldomicron{\fam=9{\mathchar"716F } }
\def\boldxi{\fam=9{\mathchar"7118 } }
\def\boldpi{\fam=9{\mathchar"7119 } }
\def\boldrho{\fam=9{\mathchar"711A } }
\def\boldsigma{\fam=9{\mathchar"711B } }
\def\boldtau{\fam=9{\mathchar"711C } }
\def\boldupsilon{\fam=9{\mathchar"711D } }
\def\boldphi{\fam=9{\mathchar"711E } }
\def\boldchi{\fam=9{\mathchar"711F } }
\def\boldpsi{\fam=9{\mathchar"7120 } }
\def\boldomega{\fam=9{\mathchar"7121 } }
\def\boldvarepsilon{\fam=9{\mathchar"7122 } }
\def\boldvartheta{\fam=9{\mathchar"7123 } }
\def\boldvarpi{\fam=9{\mathchar"7124 } }
\def\boldvarrho{\fam=9{\mathchar"7125 } }
\def\boldvarsigma{\fam=9{\mathchar"7126 } }
\def\boldvarphi{\fam=9{\mathchar"7127 } }
\def\boldGamma{\fam=6{\mathchar"7000 } }
\def\boldDelta{\fam=6{\mathchar"7001 } }
\def\boldTheta{\fam=6{\mathchar"7002 } }
\def\boldLambda{\fam=6{\mathchar"7003 } }
\def\boldXi{\fam=6{\mathchar"7004 } }
\def\boldPi{\fam=6{\mathchar"7005 } }
\def\boldSigma{\fam=6{\mathchar"7006 } }
\def\boldUpsilon{\fam=6{\mathchar"7007 } }
\def\boldPhi{\fam=6{\mathchar"7008 } }
\def\boldPsi{\fam=6{\mathchar"7009 } }
\def\boldOmega{\fam=6{\mathchar"700A } }
\def\boldmitOmega{\fam=9{\mathchar"700A } }
\def\boldmitGamma{\fam=9{\mathchar"7000 } }
\def\boldmitDelta{\fam=9{\mathchar"7001 } }
\def\boldmitTheta{\fam=9{\mathchar"7002 } }
\def\boldmitLambda{\fam=9{\mathchar"7003 } }
\def\boldmitXi{\fam=9{\mathchar"7004 } }
\def\boldmitPi{\fam=9{\mathchar"7005 } }
\def\boldmitSigma{\fam=9{\mathchar"7006 } }
\def\boldmitUpsilon{\fam=9{\mathchar"7007 } }
\def\boldmitPhi{\fam=9{\mathchar"7008 } }
\def\boldmitPsi{\fam=9{\mathchar"7009 } }
\def\boldmitOmega{\fam=9{\mathchar"700A } }

\relax

\def\double{
        \renewcommand{\baselinestretch}{2}
        \large
        \normalsize
        }

\def\single {
                \renewcommand{\baselinestretch}{1}
                \large
                \normalsize
                }

\def\etal{\hbox{\it et al.}}
\def\eg{\hbox{\it e.g.}}
\def\ie{\hbox{\it i.e.}}
\def\viz{\hbox{\it viz.}}

\def\blank#1#2#3{                                  {\bf #1} (19#3) #2}
\def\ar#1#2#3{     {\it Ann.~Rev.~Nucl.~Part.~Sci.}{\bf #1} (19#3) #2}
\def\jmp#1#2#3{    {\it J.~Math.~Phys.}            {\bf #1} (19#3) #2}
\def\np#1#2#3{     {\it Nucl.~Phys.}               {\bf #1} (19#3) #2}
\def\pl#1#2#3{     {\it Phys.~Lett.}               {\bf #1} (19#3) #2}
\def\pr#1#2#3{     {\it Phys.~Rev.}                {\bf #1} (19#3) #2}
\def\prep#1#2#3{   {\it Phys.~Rep.}                {\bf #1} (19#3) #2}
\def\prl#1#2#3{    {\it Phys.~Rev.~Lett.}          {\bf #1} (19#3) #2}
\def\rmp#1#2#3{    {\it Rev.~Mod.~Phys.}           {\bf #1} (19#3) #2}
\def\zp#1#2#3{     {\it Zeit.~fur Physik }         {\bf #1} (19#3) #2}

\def \mult{\small \otimes}
\def \plus{\small \oplus}
\def \bar{\overline}
\def \inner{\; {\LARGE \bf \cdot} \;}
\def \outer{\times}
\def \arrow{\rightarrow}
\def \doublearrow{\leftrightarrow}

\newfont{\bff}{cmtcsc10 scaled \magstep 1}

\def \oldstr {1.5}   
\def \myfootnote#1 {\footnote{#1}}
\def \chapterheading#1#2{\renewcommand{\theequation}{#1.\arabic{equation}}
                         \renewcommand{\thesection}{#1.\arabic{section}}
                         \hbox{ }
                         \vskip 1in
                         \noindent
                         \begin{tabular}{p{.5in}p{5in}}
                           {\LARGE \bf #1} & {\begin{raggedright}
                                              \LARGE \bf #2
                                              \end{raggedright}}
                         \end{tabular}
                         \vskip .5in}
\def \tablecaption#1#2{\begin{center}
                       \begin{tabular}{p{1in}p{4.5in}}
                       Table #1:  & #2
                       \end{tabular}
                       \end{center}
                       \vskip .2in}
\def \figurecaption#1#2{\renewcommand{\oldstr}{\baselinestretch}  
                        \renewcommand{\baselinestretch}{1.2}
                        \begin{center}
                        \begin{tabular}{p{1in}p{4.5in}}
                        Figure #1:  & #2
                        \end{tabular}
                        \end{center}
                        \vskip .2in
                        \renewcommand{\baselinestretch}{\oldstr}}
\def \tablist#1#2#3{\begin{tabular}{p{1in}p{4in}r} Table #1 & #2 &
                    \hskip .4231in #3 \end{tabular} \\}
\def \figlist#1#2#3{\begin{tabular}{p{1in}p{4in}r} Figure #1 & #2 &
                    \hskip .4231in #3 \end{tabular} \\}
\def \spacebetweentables{\vskip .3in}
\def \spacebetweensubtables{\vskip .2in}
\def \barcol#1{\multicolumn{1}{r|}{#1}}
\def \baselinestretch{1.2}
\def \tabcolsep{3.5pt}
\def \explanatory#1#2 {\parbox{1.5in}{#1} \hfill \parbox{4.5in}{#2}}  

\def \sign {\hbox{sign}}

\begin{document}
\begin{titlepage}
\begin{center}
February 16, 1995     \hfill    \begin{tabular}{l}
                                    LBL-36659 \\
                                    UCB-PTH-95/01 \\
                                    nucl-th/9502037 \\
                                \end{tabular}

\vskip .5in

{\large \bf Tables of $SU(3)$ Isoscalar Factors}\myfootnote{This work
was supported by the Director, Office of Energy
Research, Office of High Energy and Nuclear Physics, Division of High
Energy Physics of the U.S. Department of Energy under Contract
DE-AC03-76SF00098.}
\vskip .50in

\vskip .5in
Thomas A.~Kaeding\myfootnote{Electronic address:  {\tt kaedin@theorm.lbl.gov}.}

{\em Theoretical Physics Group\\
     Lawrence Berkeley Laboratory\\
     University of California\\
     Berkeley, California  94720}
\end{center}

\vskip .5in

\begin{abstract}
The Clebsch-Gordan coefficients of $SU(3)$ are useful in calculations
involving baryons and mesons, as well as in calculations involving
arbitrary numbers of quarks.
For the latter case, one needs the coupling constants between states
of nonintegral hypercharges.
The existing published tables are insufficient for many such applications,
and therefore we have compiled this collection.
This report supplies the isoscalar factors required to reconstruct the
Clebsch-Gordan coefficients for a large set of products of representations.
\end{abstract}
\end{titlepage}

\renewcommand{\thepage}{\roman{page}}
\setcounter{page}{2}
\mbox{ }

\vskip 1in

\begin{center}
{\bf Disclaimer}
\end{center}

\vskip .2in

\begin{scriptsize}
\begin{quotation}
This document was prepared as an account of work sponsored by the United
States Government. While this document is believed to contain correct
information, neither the United States Government nor any agency
thereof, nor The Regents of the University of California, nor any of their
employees, makes any warranty, express or implied, or assumes any legal
liability or responsibility for the accuracy, completeness, or usefulness
of any information, apparatus, product, or process disclosed, or represents
that its use would not infringe privately owned rights.  Reference herein
to any specific commercial products process, or service by its trade name,
trademark, manufacturer, or otherwise, does not necessarily constitute or
imply its endorsement, recommendation, or favoring by the United States
Government or any agency thereof, or The Regents of the University of
California.  The views and opinions of authors expressed herein do not
necessarily state or reflect those of the United States Government or any
agency thereof, or The Regents of the University of California.
\end{quotation}
\end{scriptsize}

\vskip 2in

\begin{center}
\begin{small}
{\it Lawrence Berkeley Laboratory is an equal opportunity employer.}
\end{small}
\end{center}

\newpage
\renewcommand{\baselinestretch}{1.5}
\renewcommand{\thepage}{\arabic{page}}
\setcounter{page}{1}

\noindent
Tables of $SU(3)$ Clebsch-Gordan coefficients and their isoscalar
factors\footnote{The Clebsch-Gordan coefficients are also called
{\it vector coupling
coefficients} or {\it Wigner coefficients} in the literature;  the isoscalar
factors are also called {\it Racah coefficients}.}
have been compiled in the past \cite{deswart} \cite{chilton} \cite{wali}
\cite{hecht}
\cite{rashid}, and programs have been distributed that calculate these
coefficients \cite{Akiyama}
\cite{Williamsunp} \cite{Williamspub}.
However, for calculations involving large or arbitrary numbers of quarks,
having additional tables is convenient.
For this purpose the present work extends
the available set of tables of coefficients.
Because the tables of Clebsch-Gordan coefficients would occupy far too
much space, we have extracted the isoscalar factors and present them here.
Anyone wishing a good theoretical background is suggested to consult
\cite{deswart} \cite{Carruthers} \cite{Georgi} \cite{bied}.

A few things must be discussed.
Our phase conventions are explained in Section 1.
In Section 2 we show the notation of the tables and explain how to
reconstruct the Clebsch-Gordan coefficients from the isoscalar factors.
There we also present some symmetry relations that allow us to omit
some tables.
Finally, the tables themselves are presented.

The tables of Clebsch-Gordan tables that were reduced to the present
set of isoscalar tables were generated by computer.
Most of the routines are described in \cite{cpc}.
Because we take the simple-minded approach of explicitly constructing the
representations, and use integers for exact precision, the main limitations
of the routines are due to memory allocation and integer overflows.
Any omissions in the set of isoscalar tables should be attributed to
to our lack of computing power.

\section{Conventions}

The representations of $SU(3)$ can be thought of as consisting of $SU(2)$
multiplets (henceforth called isomultiplets), each at a specific hypercharge.
We have adopted the Condon-Shortley phase convention \cite{condon}
for these isomultiplets.
This means that the eigenvalues of the isospin-raising and -lowering operators
(the $T$-spin operators) are real and positive.
What remains is to specify the relative phases between the isomultiplets
of a given representation and the overall phases of representations
in the Clebsch-Gordan series.

For the relative phases between isomultiplets in a given representation,
we have adopted the de Swart phase convention \cite{deswart}.
It corresponds to requiring that the eigenvalues of the $V$-spin operators
be real and positive.
This is simply an extension of the Condon-Shortley convention to the
$V$-spin operators.
(Recall that in flavor $SU(3)$ the $V$-spin operators interchange
$u$ and $s$ quarks.)
It is not possible to simultaneously require that the eigenvalues
of all operators be positive.

In a few cases, there is an arbitrary choice in the construction
of representations that are multiply degenerate.
By this we mean that two or more of the same representation occur
in the same Clebsch-Gordan series.
In these cases, we follow the prescription of \cite{Williamsunp}.
The highest outer degeneracy in this work is two,
and is only present when one of the factors is an octet.
The prescription of \cite{Williamsunp} corresponds to
constructing the representations in the product so that the highest-isospin
state of only one of them to couples to the isospin-1
multiplet in the factor octet.

It remains to specify the overall phases of representations in the
decomposition of the product of two irreducible representations.
Representations are named as in \cite{slansky}, with the addition
of {\bf 80} = (7, 1), {\bf 81} = (5, 2), and {\bf 90} = (4, 3)
in the usual ($p$, $q$) notation.
Throughout this work, $Y$ and $y$ denote hypercharge, $I$ and $i$
denote isospin.
We choose to follow the phase convention of de Swart \cite{deswart}.
For each representation {\bf R} in the product, consider the state with highest
third component of isospin.
Call this state the state of highest weight, and call its quantum numbers
$Y_h$, $I_h$, and $I_{h3}$.
Next consider the state in the first factor representation ({\bf r}) that has
highest isospin and couples to the highest-weight state of the product
representation;  call its quantum numbers $y_h$, $i_h$, and $i_{h3}$.
Now consider the state in the second factor representation ({\bf r$'$})
with highest isospin that couples the above two states;  its quantum
numbers are labelled $y'_h$, $i'_h$, and $i'_{h3}$.
The phase convention requires that the Clebsch-Gordan coefficient
between these three states be positive (and real):
  \begin{equation}
    \left< {\bf R} \, Y_h \, I_h \, I_{h3} |
           {\bf r} \, y_h \, i_h \, i_{h3} \,
           {\bf r'} \, y'_h \, i'_h \, i'_{h3} \right>
        \; > \; 0.
  \end{equation}
With these phase conventions,
the Clebsch-Gordan coefficients and isoscalar factors are real.

\section{Reconstruction of Clebsch-Gordan Coefficients from Isoscalar Factors}

The isoscalar factors depend of the identity of the representations,
and on the hypercharges and isospins of the isomultiplets that are coupled.
We will denote them by $F({\bf R}, Y, I; {\bf r}, y, i,$ ${\bf r'}, y', i')$.
The $SU(3)$ Clebsch-Gordan coefficients are found as products of
isoscalar factors and $SU(2)$ Clebsch-Gordan coefficients:
  \begin{equation}
    \left< {\bf R} \, Y \, I \, I_3 | {\bf r} \, y \, i \, i_3 \,
                                      {\bf r'} \, y' \, i' \, i'_3 \right>
        = F({\bf R}, Y, I; {\bf r}, y, i, {\bf r'}, y', i')
        \times \left< I \, I_3 | i \, i_3 \, i' \, i'_3 \right>.
  \end{equation}
The $SU(2)$ tables can be reconstructed from Tables $1^3$, $2^3$, $3^3$,
and $4^3$ of \cite{condon}.
For isospin less than or equal to two, they can conveniently be
found in the {\it Review of Particle Properties} \cite{pdg}.
Note the easily overlooked relation
  \begin{equation}
    \left< I \, I_3 | i \, i_3 \, i' \, i'_3 \right> = (-1)^{I - i - i'}
        \left< I \, I_3 | i' \, i'_3 \, i \, i_3 \right>.
  \label{switchisos}
  \end{equation}

There are two symmetry relations among the isoscalar factors that will allow
us to omit many tables from our exposition.
Those tables can be reconstructed from those that are present, with the
help of the phase factors involved in these symmetry relations.
Both relations come from \cite{deswart}, but we rewrite them in our
notation.\myfootnote{Our $\xi$ is the $\xi_1$ of \cite{deswart};  $\zeta$ is
$\xi_3$ of \cite{deswart}.}
The first involves the order of the factor representations.
If the order is reversed, then a phase $\xi$ may enter:
  \begin{equation}
    F({\bf R}, Y, I; {\bf r'}, y', i', {\bf r}, y, i)
        = (-1)^{I - i - i'} \xi ({\bf R}; {\bf r}, {\bf r'})
          F({\bf R}, Y, I; {\bf r}, y, i, {\bf r'}, y', i').
  \label{defxi}
  \end{equation}
The factor $(-1)^{I - i - i'}$ comes from Equation \ref{switchisos}.
The phase $\xi ({\bf R}; {\bf r}, {\bf r'})$
does not depend on the quantum numbers of the states,
but only on the identity of the representations ${\bf r}$, ${\bf r'}$,
and ${\bf R}$, and on the phase conventions described in the previous section.
The second symmetry relation involves the conjugation of the representations:
  \begin{equation}
    F(\bar{\bf R}, Y, I; \bar{\bf r}, y, i, \bar{\bf r'}, y', i')
        = (-1)^{I - i - i'} \zeta ({\bf R}; {\bf r}, {\bf r'})
          F({\bf R}, \hbox{-}Y, I; {\bf r}, \hbox{-}y, i,
                                   {\bf r'}, \hbox{-}y', i').
  \label{defzeta}
  \end{equation}
Here $\zeta ({\bf R}; {\bf r}, {\bf r'})$
also does not depend on the quantum numbers of the states involved, but
only on the identities of the representations and on our phase conventions.
For these relations, we naturally define $\bar{\bf 1}$ $\equiv$ {\bf 1},
$\bar{\bf 8}$ $\equiv$ {\bf 8},
$\bar{\bf 27}$ $\equiv$ {\bf 27}, and $\bar{\bf 64}$ $\equiv$ {\bf 64}.
It is easy to show from Equations \ref{defxi} and \ref{defzeta} that
  \begin{equation}
    \xi (\bar{\bf R}; \bar{\bf r}, \bar{\bf r'})
        = \xi ({\bf R}; {\bf r}, {\bf r'}).
  \label{barxi}
  \end{equation}
Then it can be shown (using Equation \ref{barxi}) that
  \begin{equation}
    \zeta ({\bf R}; {\bf r'}, {\bf r}) = \zeta ({\bf R}; {\bf r}, {\bf r'}).
  \end{equation}
The $\xi$ and $\zeta$ needed to construct the omitted tables are presented
in Tables 2 and 3.
We should note that it is not necessary to construct the tables related by
Equations \ref{defxi} and \ref{defzeta} in order to find the phase factors.
They are found by considering the highest-isospin states of the
representations in the product.
Suppose that such a state and the highest-isospin states coupling to it
are described as in Section 1.
Then from Equation \ref{switchisos} we find simply that
  \begin{equation}
    \zeta ({\bf R}; {\bf r}, {\bf r'})
        = (-1)^{I_h - i_h - i'_h}.
  \end{equation}
If in the reversed product ${\bf r'}$ $\mult$ {\bf r} the highest-isospin state
in ${\bf r'}$ that couples to $I_h$ in {\bf R} has quantum numbers
$y^{\prime rev}_h$ and $i^{\prime rev}_h$, and the highest isospin in
{\bf r} that couples to these two has $y^{rev}_h$ and $i^{rev}_h$.
Then
  \begin{equation}
    \xi ({\bf R}; {\bf r}, {\bf r'})
        = (-1)^{I_h - i^{rev}_h - i^{\prime rev}_h}
        \times \sign \left[ F ({\bf R}, Y_h, I_h;
                                 {\bf r}, y^{rev}_h, i^{rev}_h,
                                 {\bf r'}, y^{\prime rev}_h,
                                           i^{\prime rev}_h) \right],
  \end{equation}
where $\sign(x)$ = $x / |x|$.

\section{Tables}

The tables follow.
Table 1 is a list of the products of representations whose isoscalar
factors are given in this work.
Tables 2 and 3 give the phase factors which are described in the previous
section and which can be used to generate other tables from the ones presented.
The tables of isoscalar factors appear last.
Tables for {\bf 8} $\mult$ {\bf 8}, {\bf 10} $\mult$ {\bf 8}, {\bf 10} $\mult$
{\bf 10}, $\bar{\bf 10}$ $\mult$ {\bf 10}, and {\bf 27} $\mult$ {\bf 8}
appear in \cite{deswart} and \cite{chilton};  they are not reproduced here.

In the tables, a square root is assumed to appear over each entry
(signs are outside the square roots).
Thus,
  \begin{equation}
\begin{array}{rrrr|c|}
    &&&& {\bf R} \\
    \multicolumn{4}{c|}{{\bf r} \; {\small \otimes} \; {\bf r'}} & Y \\
    &&&& I \\
  \hline
    y & i & y' & i' & \pm C \\
  \hline
  \end{array}
  \end{equation}
means that the isoscalar factor
  \begin{equation}
    F({\bf R}, Y, I; {\bf r}, y, i, {\bf r'}, y', i')
        = \pm \sqrt{C}.
  \end{equation}

\section*{Acknowledgements}

We wish to thank J.~J.~de Swart for his explanation of his phase conventions
and for his hospitality.

This work was supported by the Director, Office of Energy
Research, Office of High Energy and Nuclear Physics, Division of High
Energy Physics of the U.S. Department of Energy under Contract
DE-AC03-76SF00098.

\newpage
\tablecaption{1}{List of tables of isoscalar factors that follow.}
\begin{center}
\begin{tabular}{lrrclcp{3in}}
  Table &  4   & \quad {\bf 3}          & $\mult$ & {\bf 3}
              & = & {\bf 6} $\plus$ $\bar{\bf 3}$ \\
  Table &  5   &  $\bar{\bf 3}$ & $\mult$ & {\bf 3}
              & = & {\bf 8} $\plus$ {\bf 1} \\
  Table &  6   &  {\bf 6}          & $\mult$ & {\bf 3}
              & = & {\bf 10} $\plus$ {\bf 8} \\
  Table &  7   &  {\bf 6}          & $\mult$ & $\bar{{\bf 3}}$
              & = & {\bf 15} $\plus$ {\bf 3} \\
  Table &  8   &  {\bf 6}          & $\mult$ & {\bf 6}
              & = & {\bf 15$'$} $\plus$ {\bf 15} $\plus$ $\bar{\bf 6}$ \\
  Table &  9   &  $\bar{{\bf 6}}$  & $\mult$ & {\bf 6}
              & = & {\bf 27} $\plus$ {\bf 8} $\plus$ {\bf 1} \\
  Table &  10  &  {\bf 8}          & $\mult$ & {\bf 3}
              & = & {\bf 15} $\plus$ $\bar{\bf 6}$ $\plus$ {\bf 3} \\
  Table &  11  &  {\bf 8}          & $\mult$ & {\bf 6}
              & = & $\bar{\bf 24}$ $\plus$ $\bar{\bf 15}$ $\plus$ {\bf 6}
                       $\plus$ $\bar{\bf 3}$ \\
\end{tabular}
\end{center}

\spacebetweensubtables
\begin{center}
\begin{tabular}{lrrclcp{3in}}
  Table &  29  & \quad $\bar{\bf 21}$ & $\mult$ & {\bf 3}
               & = & {\bf 28} $\plus$ {\bf 35} \\
  Table &  30  & $\bar{\bf 21}$ & $\mult$ & $\bar{\bf 3}$
               & = & {\bf 48} $\plus$ {\bf 15$'$} \\
  Table &  31  & $\bar{\bf 21}$ & $\mult$ & {\bf 6}
               & = & {\bf 36} $\plus$ {\bf 48} $\plus$ {\bf 42} \\
  Table &  32  & $\bar{\bf 21}$ & $\mult$ & $\bar{\bf 6}$
               & = & {\bf 81} $\plus$ {\bf 35} $\plus$ {\bf 10} \\
  Table &  33  & $\bar{\bf 21}$ & $\mult$ & {\bf 8}
               & = & $\bar{\bf 63}$ $\plus$ $\bar{\bf 60}$ $\plus$
                          $\bar{\bf 21}$ $\plus$ $\bar{\bf 24}$ \\
  Table &  34  & $\bar{\bf 21}$ & $\mult$ & {\bf 10}
               & = & $\bar{\bf 45}$ $\plus$ $\bar{\bf 63}$ $\plus$
                          $\bar{\bf 60}$ $\plus$ $\bar{\bf 42}$ \\
  Table &  35  & $\bar{\bf 24}$ & $\mult$ & {\bf 3}
               & = & {\bf 35} $\plus$ {\bf 27} $\plus$ {\bf 10} \\
  Table &  36  & $\bar{\bf 24}$ & $\mult$ & $\bar{\bf 3}$
               & = & {\bf 42} $\plus$ {\bf 15$'$} $\plus$ {\bf 15} \\
\end{tabular}
\end{center}

\newpage
\tablecaption{2}{Phase factors $\xi$.
                 Here ($+$) denotes $+1$ and ($-$) denotes $-1$.}
\begin{center}
\begin{tabular}{cc|ccccccccc|}
  \multicolumn{2}{c|}{$\xi$} & \multicolumn{9}{c|}{${\bf R}$} \\
\hline
   ${\bf r}$ & ${\bf r'}$ &
      {\bf 1} & {\bf 8} & {\bf 10} & $\bar{\bf 10}$ &
      {\bf 27} & {\bf 35} &
      {\bf 28} & {\bf 81} & {\bf 80} \\
\hline
  $\bar{\bf 3}$ & {\bf 3}   & $-$ & $+$ &&&&&&& \\
  {\bf 6} & {\bf 3}         && $-$ & $+$ &&&&&& \\
  $\bar{\bf 6}$ & {\bf 6}   & $+$ & $-$ &&& $+$ &&&& \\
  {\bf 15} & $\bar{\bf 3}$  && $-$ & $-$ && $+$ &&&& \\
  {\bf 15} & {\bf 6}        && $+$ & $-$ & $+$ & $-$ & $+$ &&& \\
  {\bf 15$'$} & $\bar{\bf 3}$     &&& $-$ &&& $+$ &&& \\
  {\bf 15$'$} & {\bf 6}           &&&&& $+$ & $-$ & $+$ && \\
  $\bar{\bf 21}$ & {\bf 3}        &&&&&& $-$ & $+$ && \\
  $\bar{\bf 21}$ & $\bar{\bf 6}$  &&& $+$ &&& $-$ && $+$ & \\
  $\bar{\bf 24}$ & {\bf 3}        &&& $-$ && $-$ & $+$ &&& \\
  {\bf 36} & $\bar{\bf 3}$        &&&&&&& $-$ && $+$ \\
\hline
\end{tabular}

\spacebetweensubtables
\begin{tabular}{cc|ccccccccc|}
  \multicolumn{2}{c|}{$\xi$} & \multicolumn{9}{c|}{${\bf R}$} \\
\hline
   ${\bf r}$ & ${\bf r'}$ &
      {\bf 3} & $\bar{\bf 6}$ & {\bf 15}$_1$ & {\bf 15}$_2$ &
      {\bf 15$'$} & {\bf 24} &
      {\bf 42} & {\bf 48} & {\bf 36} \\
\hline
  {\bf 6} & $\bar{\bf 3}$  & $-$ && $+$ &&&&&& \\
  {\bf 6} & {\bf 6}        && $+$ & $-$ && $+$ &&&& \\
  {\bf 8} & {\bf 3}        & $-$ & $-$ & $+$ &&&&&& \\
  {\bf 10} & {\bf 3}       &&& $-$ && $+$ &&&& \\
  {\bf 10} & $\bar{\bf 6}$ & $+$ && $-$ &&&& $+$ && \\
  {\bf 15} & {\bf 8}         & $+$ & $+$ & $+$ & $-$ & $-$ & $-$ & $+$ && \\
  {\bf 15} & {\bf 10}        && $-$ & $+$ && $-$ & $+$ & $-$ & $+$ & \\
  {\bf 15$'$} & {\bf 8}         &&& $+$ && $-$ && $-$ & $+$ & \\
  {\bf 15$'$} & {\bf 10}        &&&&&& $-$ & $+$ & $-$ & $+$ \\
  $\bar{\bf 21}$ & $\bar{\bf 3}$ &&&&& $-$ &&& $+$ & \\
  $\bar{\bf 21}$ & {\bf 6}       &&&&&&& $+$ & $-$ & $+$ \\
  $\bar{\bf 24}$ & $\bar{\bf 3}$ &&& $-$ && $-$ && $+$ && \\
  $\bar{\bf 24}$ & {\bf 6}       &&& $+$ && $-$ & $+$ & $-$ & $+$ & \\
  {\bf 27} & {\bf 3}             &&& $-$ &&& $-$ & $+$ && \\
  {\bf 28} & {\bf 3}             &&&&&&&& $-$ & $+$ \\
  {\bf 35} & {\bf 3}             &&&&& $-$ && $-$ & $+$ & \\
\hline
\end{tabular}

\spacebetweensubtables
\begin{tabular}{cc|ccccccccccc|}
  \multicolumn{2}{c|}{$\xi$} & \multicolumn{11}{c|}{${\bf R}$} \\
\hline
   ${\bf r}$ & ${\bf r'}$ &
      $\bar{\bf 3}$ & {\bf 6} & $\bar{\bf 15}$ &
      $\bar{\bf 15'}$ & $\bar{\bf 24}_1$ &  $\bar{\bf 24}_2$ & $\bar{\bf 21}$ &
      $\bar{\bf 42}$ & $\bar{\bf 60}$ & $\bar{\bf 63}$ & $\bar{\bf 45}$ \\
\hline
  {\bf 3} & {\bf 3}        & $-$ & $+$ &&&&&&&&& \\
  {\bf 8} & {\bf 6}        & $+$ & $-$ & $-$ && $+$ &&&&&& \\
  {\bf 10} & $\bar{\bf 3}$ && $-$ &&& $+$ &&&&&& \\
  {\bf 10} & {\bf 6}       &&& $+$ && $-$ && $+$ &&&& \\
  {\bf 15} & {\bf 3}       && $-$ & $-$ && $+$ &&&&&& \\
  {\bf 15} & $\bar{\bf 6}$ & $+$ & $+$ & $-$ && $-$ &&& $+$ &&& \\
  {\bf 15$'$} & {\bf 3}       &&&&& $-$ && $+$ &&&& \\
  {\bf 15$'$} & $\bar{\bf 6}$ && $+$ &&& $-$ &&&& $+$ && \\
  {\bf 15$'$} & {\bf 15$'$}   &&&& $+$ &&&& $-$ & $+$ & $-$ & $+$ \\
  $\bar{\bf 21}$ & {\bf 8}    &&&&& $+$ && $-$ && $-$ & $+$ & \\
  $\bar{\bf 21}$ & {\bf 10}   &&&&&&&& $-$ & $+$ & $-$ & $+$ \\
  $\bar{\bf 24}$ & {\bf 8}    && $+$ & $+$ && $+$ & $-$ & $-$ & $-$ & $+$ &&\\
  {\bf 27} & {\bf 6}          && $+$ & $+$ & $+$ & $-$ &&& $-$ & $+$ && \\
  {\bf 28} & $\bar{\bf 3}$    &&&&&&& $-$ &&& $+$ & \\
  {\bf 28} & {\bf 6}          &&&&&&&&& $+$ & $-$ & $+$ \\
  {\bf 35} & $\bar{\bf 3}$    &&&&& $-$ && $-$ && $+$ && \\
  {\bf 35} & {\bf 6}          &&&&& $+$ && $-$ & $+$ & $-$ & $+$ & \\
  {\bf 36} & {\bf 3}          &&&&&&&&&& $-$ & $+$ \\
\hline
\end{tabular}
\end{center}

\newpage
\tablecaption{3}{Phase factors $\zeta$.
                 Here ($+$) denotes $+1$ and ($-$) denotes $-1$.}
\begin{center}
\begin{tabular}{cc|ccccccccc|}
  \multicolumn{2}{c|}{$\zeta$} & \multicolumn{9}{c|}{${\bf R}$} \\
\hline
   ${\bf r}$ & ${\bf r'}$ &
      {\bf 1} & {\bf 8} & {\bf 10} & $\bar{\bf 10}$ &
      {\bf 27} & {\bf 35} &
      {\bf 28} & {\bf 81} & {\bf 80} \\
\hline
  $\bar{\bf 3}$ & {\bf 3}   & $-$ & $+$ &&&&&&& \\
  {\bf 6} & {\bf 3}         && $+$ & $+$ &&&&&& \\
  $\bar{\bf 6}$ & {\bf 6}   & $+$ & $-$ &&& $+$ &&&& \\
  {\bf 15} & $\bar{\bf 3}$  && $-$ & $+$ && $+$ &&&& \\
  {\bf 15} & {\bf 6}        && $-$ & $-$ & $+$ & $+$ & $+$ &&& \\
  {\bf 15$'$} & $\bar{\bf 3}$     &&& $-$ &&& $+$ &&& \\
  {\bf 15$'$} & {\bf 6}           &&&&& $+$ & $+$ & $+$ && \\
  $\bar{\bf 21}$ & {\bf 3}        &&&&&& $+$ & $+$ && \\
  $\bar{\bf 21}$ & $\bar{\bf 6}$  &&& $+$ &&& $-$ && $+$ & \\
  $\bar{\bf 24}$ & {\bf 3}        &&& $-$ && $+$ & $+$ &&& \\
  {\bf 36} & $\bar{\bf 3}$        &&&&&&& $-$ && $+$ \\
\hline
\end{tabular}

\spacebetweensubtables
\begin{tabular}{cc|ccccccccc|}
  \multicolumn{2}{c|}{$\zeta$} & \multicolumn{9}{c|}{${\bf R}$} \\
\hline
   ${\bf r}$ & ${\bf r'}$ &
      {\bf 3} & $\bar{\bf 6}$ & {\bf 15}$_1$ & {\bf 15}$_2$ &
      {\bf 15$'$} & {\bf 24} &
      {\bf 42} & {\bf 48} & {\bf 36} \\
\hline
  {\bf 6} & $\bar{\bf 3}$  & $-$ && $+$ &&&&&& \\
  {\bf 6} & {\bf 6}        && $+$ & $+$ && $+$ &&&& \\
  {\bf 8} & {\bf 3}        & $-$ & $+$ & $+$ &&&&&& \\
  {\bf 10} & {\bf 3}       &&& $+$ && $+$ &&&& \\
  {\bf 10} & $\bar{\bf 6}$ & $+$ && $-$ &&&& $+$ && \\
  {\bf 15} & {\bf 8}         & $+$ & $-$ & $+$ & $-$ & $+$ & $+$ & $+$ && \\
  {\bf 15} & {\bf 10}        && $-$ & $-$ && $-$ & $+$ & $+$ & $+$ & \\
  {\bf 15$'$} & {\bf 8}         &&& $-$ && $-$ && $+$ & $+$ & \\
  {\bf 15$'$} & {\bf 10}        &&&&&& $+$ & $+$ & $+$ & $+$ \\
  $\bar{\bf 21}$ & $\bar{\bf 3}$ &&&&& $-$ &&& $+$ & \\
  $\bar{\bf 21}$ & {\bf 6}       &&&&&&& $+$ & $+$ & $+$ \\
  $\bar{\bf 24}$ & $\bar{\bf 3}$ &&& $-$ && $+$ && $+$ && \\
  $\bar{\bf 24}$ & {\bf 6}       &&& $-$ && $-$ & $+$ & $+$ & $+$ & \\
  {\bf 28} & {\bf 3}             &&&&&&&& $+$ & $+$ \\
  {\bf 27} & {\bf 3}             &&& $-$ &&& $+$ & $+$ && \\
  {\bf 35} & {\bf 3}             &&&&& $-$ && $+$ & $+$ & \\
\hline
\end{tabular}

\spacebetweensubtables
\begin{tabular}{cc|ccccccccccc|}
  \multicolumn{2}{c|}{$\zeta$} & \multicolumn{11}{c|}{${\bf R}$} \\
\hline
   ${\bf r}$ & ${\bf r'}$ &
      $\bar{\bf 3}$ & {\bf 6} & $\bar{\bf 15}_1$ &
      $\bar{\bf 15'}$ & $\bar{\bf 24}_1$ &  $\bar{\bf 24}_2$ & $\bar{\bf 21}$ &
      $\bar{\bf 42}$ & $\bar{\bf 60}$ & $\bar{\bf 63}$ & $\bar{\bf 45}$ \\
\hline
  {\bf 3} & {\bf 3}        & $+$ & $+$ &&&&&&&&& \\
  {\bf 8} & {\bf 6}        & $-$ & $-$ & $+$ && $+$ &&&&&& \\
  {\bf 10} & $\bar{\bf 3}$ && $-$ &&& $+$ &&&&&& \\
  {\bf 10} & {\bf 6}       &&& $+$ && $+$ && $+$ &&&& \\
  {\bf 15} & {\bf 3}       && $-$ & $+$ && $+$ &&&&&& \\
  {\bf 15} & $\bar{\bf 6}$ & $+$ & $-$ & $-$ && $+$ &&& $+$ &&& \\
  {\bf 15$'$} & {\bf 3}       && $-$ & $+$ && $+$ &&&&&& \\
  {\bf 15$'$} & $\bar{\bf 6}$ & $+$ & $-$ & $-$ && $+$ &&& $+$ &&& \\
  {\bf 15$'$} & {\bf 15$'$}   &&&& $+$ &&&& $+$ & $+$ & $+$ & $+$ \\
  $\bar{\bf 21}$ & {\bf 8}    &&&&& $-$ && $-$ && $+$ & $+$ & \\
  $\bar{\bf 21}$ & {\bf 10}   &&&&&&&& $+$ & $+$ & $+$ & $+$ \\
  $\bar{\bf 24}$ & {\bf 8}    && $+$ & $-$ && $+$ & $-$ & $+$ & $+$ & $+$ &&\\
  {\bf 27} & {\bf 6}          && $+$ & $-$ & $+$ & $-$ &&& $+$ & $+$ && \\
  {\bf 28} & $\bar{\bf 3}$    &&&&&&& $-$ &&& $+$ & \\
  {\bf 28} & {\bf 6}          &&&&&&&&& $+$ & $+$ & $+$ \\
  {\bf 35} & $\bar{\bf 3}$    &&&&& $-$ && $+$ && $+$ && \\
  {\bf 35} & {\bf 6}          &&&&& $+$ && $-$ & $+$ & $+$ & $+$ & \\
  {\bf 36} & {\bf 3}          &&&&&&&&&& $+$ & $+$ \\
\hline
\end{tabular}
\end{center}

\newpage


\spacebetweentables
\tablecaption{4}{Isoscalar factors for {\bf 3} $\mult$ {\bf 3}.
                   Notation for this and following tables
                   is explained in the text.
                   In these tables a square root is assumed over each entry.}
\begin{center}
\begin{tabular}{rrrr|rrrrr|}
 &&&& $\bar{\bf 3}$ & {\bf 6} & $\bar{\bf 3}$ & {\bf 6} & {\bf 6} \\
 \multicolumn{4}{c|}{{\bf 3} $\mult$ {\bf 3}} & $\frac{2}{3}$ & $\frac{2}{3}$ &
         -$\frac{1}{3}$ & -$\frac{1}{3}$ & -$\frac{4}{3}$ \\
 &&&& 0 & 1 & $\frac{1}{2}$ & $\frac{1}{2}$ & 0 \\
\hline
 $\frac{1}{3}$ & $\frac{1}{2}$ & $\frac{1}{3}$ & $\frac{1}{2}$ &
      \multicolumn{1}{r|}{-1} & \multicolumn{1}{r|}{1} &&& \\
\cline{5-8}
 $\frac{1}{3}$ & $\frac{1}{2}$ & -$\frac{2}{3}$ & 0 &
      & \multicolumn{1}{r|}{} & $\frac{1}{2}$ &
      \multicolumn{1}{r|}{$\frac{1}{2}$} & \\
 -$\frac{2}{3}$ & 0 & $\frac{1}{3}$ & $\frac{1}{2}$ &
      & \multicolumn{1}{r|}{} & -$\frac{1}{2}$ &
      \multicolumn{1}{r|}{$\frac{1}{2}$} & \\
\cline{7-9}
 -$\frac{2}{3}$ & 0 & -$\frac{2}{3}$ & 0 &
      &&& \multicolumn{1}{r|}{} & 1 \\
\hline
\end{tabular}
\end{center}


\spacebetweentables
\tablecaption{5}{Isoscalar factors for $\bar{\bf 3}$ $\mult$ {\bf 3}.}
\begin{center}
\begin{tabular}{rrrr|rrrrr|}
 &&&& {\bf 8} & {\bf 1} & {\bf 8} & {\bf 8} & {\bf 8} \\
 \multicolumn{4}{c|}{$\bar{\bf 3}$ $\mult$ {\bf 3}} & 1 & 0 & 0 & 0 & -1 \\
 &&&& $\frac{1}{2}$ & 0 & 0 & 1 & $\frac{1}{2}$ \\
\hline
 $\frac{2}{3}$ & 0 & $\frac{1}{3}$ & $\frac{1}{2}$ &
      \multicolumn{1}{r|}{1} &&&& \\
\cline{5-7}
 $\frac{2}{3}$ & 0 & -$\frac{2}{3}$ & 0 &
      \multicolumn{1}{r|}{} & $\frac{1}{3}$ &
      \multicolumn{1}{r|}{$\frac{2}{3}$} && \\
\cline{8-8}
 -$\frac{1}{3}$ & $\frac{1}{2}$ & $\frac{1}{3}$ & $\frac{1}{2}$ &
      \multicolumn{1}{r|}{} & $\frac{2}{3}$ &
      \multicolumn{1}{r|}{-$\frac{1}{3}$} &
      \multicolumn{1}{r|}{1} & \\
\cline{6-9}
 -$\frac{1}{3}$ & $\frac{1}{2}$ & -$\frac{2}{3}$ & 0 &
      &&& \multicolumn{1}{r|}{} & 1 \\
\hline
\end{tabular}
\end{center}


\spacebetweentables
\tablecaption{6}{Isoscalar factors for {\bf 6} $\mult$ {\bf 3}.}
\begin{center}
\begin{tabular}{rrrr|rrrrrrrr|}
 &&&& {\bf 8} & {\bf 10} & {\bf 8} & {\bf 8} & {\bf 10} & {\bf 8} &
         {\bf 10} & {\bf 10} \\
 \multicolumn{4}{c|}{{\bf 6} $\mult$ {\bf 3}} & 1 & 1 & 0 & 0 & 0 &
         -1 & -1 & -2 \\
 &&&& $\frac{1}{2}$ & $\frac{3}{2}$ & 0 & 1 & 1 & $\frac{1}{2}$ &
          $\frac{1}{2}$ & 0 \\
\hline
 $\frac{2}{3}$ & 1 & $\frac{1}{3}$ & $\frac{1}{2}$ &
    \barcol{-1} & \barcol{1} & &&&&& \\
\cline{5-9}
 -$\frac{1}{3}$ & $\frac{1}{2}$ & $\frac{1}{3}$ & $\frac{1}{2}$ &
    & \barcol{} & \barcol{-1} &
    -$\frac{1}{3}$ & \barcol{$\frac{2}{3}$} & && \\
\cline{7-7}
 $\frac{2}{3}$ & 1 & -$\frac{2}{3}$ & 0 &
    && \barcol{} & $\frac{2}{3}$ &
    \barcol{$\frac{1}{3}$} & && \\
\cline{8-11}
 -$\frac{1}{3}$ & $\frac{1}{2}$ & -$\frac{2}{3}$ & 0 &
    &&&& \barcol{} &
    $\frac{1}{3}$ & \barcol{$\frac{2}{3}$} & \\
 -$\frac{4}{3}$ & 0 & $\frac{1}{3}$ & $\frac{1}{2}$ &
    &&&& \barcol{} &
    -$\frac{2}{3}$ & \barcol{$\frac{1}{3}$} & \\
\cline{10-12}
 -$\frac{4}{3}$ & 0 & -$\frac{2}{3}$ & 0 &
    &&&&&& \barcol{} & 1 \\
\hline
\end{tabular}
\end{center}


\spacebetweentables
\tablecaption{7}{Isoscalar factors for {\bf 6} $\mult$ $\bar{\bf 3}$.}
\begin{center}
\begin{tabular}{rrrr|rrrrrrrr|}
 &&&& {\bf 15} & {\bf 3} & {\bf 15} & {\bf 15} & {\bf 3} & {\bf 15} &
         {\bf 15} & {\bf 15} \\
 \multicolumn{4}{c|}{{\bf 6} $\mult$ $\bar{\bf 3}$} &
         $\frac{4}{3}$ & $\frac{1}{3}$ & $\frac{1}{3}$ & $\frac{1}{3}$ &
         -$\frac{2}{3}$ & -$\frac{2}{3}$ & -$\frac{2}{3}$ & -$\frac{5}{3}$ \\
 &&&& 1 & $\frac{1}{2}$ & $\frac{1}{2}$ & $\frac{3}{2}$ & 0 &
        0 & 1 & $\frac{1}{2}$ \\
\hline
 $\frac{2}{3}$ & 1 & $\frac{2}{3}$ & 0 &
    \barcol{1} & & &&&&& \\
\cline{5-7}
 -$\frac{1}{3}$ & $\frac{1}{2}$ & $\frac{2}{3}$ & 0 &
    \barcol{} & -$\frac{1}{4}$ &
    \barcol{$\frac{3}{4}$} & &&&& \\
\cline{8-8}
 $\frac{2}{3}$ & 1 & -$\frac{1}{3}$ & $\frac{1}{2}$ &
    \barcol{} & $\frac{3}{4}$ &
    \barcol{$\frac{1}{4}$} & \barcol{1} & &&& \\
\cline{6-10}
 -$\frac{4}{3}$ & 0 & $\frac{2}{3}$ & 0 &
    &&& \barcol{} &
    -$\frac{1}{2}$ & \barcol{$\frac{1}{2}$} & & \\
\cline{11-11}
 -$\frac{1}{3}$ & $\frac{1}{2}$ & -$\frac{1}{3}$ & $\frac{1}{2}$ &
    &&& \barcol{} &
    $\frac{1}{2}$ & \barcol{$\frac{1}{2}$} &
    \barcol{1} & \\
\cline{9-12}
 -$\frac{4}{3}$ & 0 & -$\frac{1}{3}$ & $\frac{1}{2}$ &
    &&&&&& \barcol{} & 1 \\
\hline
\end{tabular}
\end{center}


\spacebetweentables
\tablecaption{8}{Isoscalar factors for {\bf 6} $\mult$ {\bf 6}.}
\begin{center}
\begin{tabular}{rrrr|rrrrrrrrrrrrrr|}
 & & & & $\bar{\bf 6}$ & {\bf 15} & {\bf 15$'$} & $\bar{\bf 6}$ &
           {\bf 15} & {\bf 15} & {\bf 15$'$} &
      {\bf 15} & $\bar{\bf 6}$ & {\bf 15} & {\bf 15$'$} & {\bf 15} &
         {\bf 15$'$} & {\bf 15$'$} \\
 \multicolumn{4}{c|}{{\bf 6} $\mult$ {\bf 6}} &
    $\frac{4}{3}$ & $\frac{4}{3}$ & $\frac{4}{3}$ &
    $\frac{1}{3}$ & $\frac{1}{3}$ & $\frac{1}{3}$ & $\frac{1}{3}$ &
    -$\frac{2}{3}$ & -$\frac{2}{3}$ & -$\frac{2}{3}$ &
    -$\frac{2}{3}$ & -$\frac{5}{3}$ & -$\frac{5}{3}$ & -$\frac{8}{3}$ \\
 & & & & 0 & 1 & 2 &
     $\frac{1}{2}$ & $\frac{1}{2}$ & $\frac{3}{2}$ & $\frac{3}{2}$ &
     0 & 1 & 1 & 1 & $\frac{1}{2}$ & $\frac{1}{2}$ & 0 \\
\hline
 $\frac{2}{3}$ & 1 & $\frac{2}{3}$ & 1 &
    \barcol{1} & \barcol{-1} & \barcol{1} & &&&&&&&&&& \\
\cline{5-11}
 $\frac{2}{3}$ & 1 & -$\frac{1}{3}$ & $\frac{1}{2}$ &
    && \barcol{} & -$\frac{1}{2}$ & \barcol{-$\frac{1}{2}$} &
    $\frac{1}{2}$ & \barcol{$\frac{1}{2}$} & &&&&&& \\
 -$\frac{1}{3}$ & $\frac{1}{2}$ & $\frac{2}{3}$ & 1 &
    && \barcol{} & $\frac{1}{2}$ & \barcol{-$\frac{1}{2}$} &
    -$\frac{1}{2}$ & \barcol{$\frac{1}{2}$} & &&&&&& \\
\cline{8-15}
 -$\frac{1}{3}$ & $\frac{1}{2}$ & -$\frac{1}{3}$ & $\frac{1}{2}$ &
    &&&&&& \barcol{} &
    \barcol{-1} & -$\frac{1}{3}$ & 0 & \barcol{$\frac{2}{3}$} & && \\
\cline{12-12}
 $\frac{2}{3}$ & 1 & -$\frac{4}{3}$ & 0 &
    &&&&&&&
    \barcol{} & $\frac{1}{3}$ & $\frac{1}{2}$ & \barcol{$\frac{1}{6}$} & && \\
\hline
\end{tabular}
\end{center}


\newpage
\tablecaption{9}{Isoscalar factors for $\bar{\bf 6}$ $\mult$ {\bf 6}.}
\begin{center}
\begin{tabular}{rrrr|rrrrrrrrrrrrrr|}
 & & & & {\bf 27} & {\bf 8} & {\bf 27} & {\bf 27} & {\bf 1} & {\bf 8} &
    {\bf 27} & {\bf 8} & {\bf 27} & {\bf 27} &
    {\bf 8} & {\bf 27} & {\bf 27} & {\bf 27} \\
 \multicolumn{4}{c|}{$\bar{\bf 6}$ $\mult$ {\bf 6}} &
    2 & 1 & 1 & 1 & 0 & 0 & 0 & 0 & 0 & 0 & -1 & -1 & -1 & - 2 \\
 & & & & 1 & $\frac{1}{2}$ & $\frac{1}{2}$ & $\frac{3}{2}$ & 0 & 0 & 0 &
    1 & 1 & 2 & $\frac{1}{2}$ & $\frac{1}{2}$ & $\frac{3}{2}$ & 1 \\
\hline
 $\frac{4}{3}$ & 0 & $\frac{2}{3}$ & 1 &
    \barcol{1} &&&&&&&&&&&&& \\
\cline{5-7}
 $\frac{4}{3}$ & 0 & -$\frac{1}{3}$ & $\frac{1}{2}$ &
    \barcol{} & $\frac{2}{5}$ & \barcol{$\frac{3}{5}$} & &&&&&&&&&& \\
\cline{8-8}
 $\frac{1}{3}$ & $\frac{1}{2}$ & $\frac{2}{3}$ & 1 &
    \barcol{} & $\frac{3}{5}$ & \barcol{-$\frac{2}{5}$} & \barcol{1} &
    &&&&&&&&& \\
\cline{6-11}
 $\frac{4}{3}$ & 0 & -$\frac{4}{3}$ & 0 &
    &&& \barcol{} & $\frac{1}{6}$ & $\frac{8}{15}$ & \barcol{$\frac{3}{10}$} &
    &&&&&& \\
\cline{12-13}
 $\frac{1}{3}$ & $\frac{1}{2}$ & -$\frac{1}{3}$ & $\frac{1}{2}$ &
    &&& \barcol{} & $\frac{1}{3}$ & $\frac{1}{15}$ & \barcol{-$\frac{3}{5}$} &
    $\frac{1}{5}$ & \barcol{$\frac{4}{5}$} & &&&& \\
\cline{14-14}
 -$\frac{2}{3}$ & 1 & $\frac{2}{3}$ & 1 &
    &&& \barcol{} & $\frac{1}{2}$ & -$\frac{2}{5}$ & \barcol{$\frac{1}{10}$} &
    $\frac{4}{5}$ & \barcol{-$\frac{1}{5}$} & \barcol{1} & &&& \\
\cline{9-16}
 $\frac{1}{3}$ & $\frac{1}{2}$ & -$\frac{4}{3}$ & 0 &
    &&&&&&&&& \barcol{} & $\frac{2}{5}$ & \barcol{$\frac{3}{5}$} & & \\
\cline{17-17}
 -$\frac{2}{3}$ & 1 & -$\frac{1}{3}$ & $\frac{1}{2}$ &
    &&&&&&&&& \barcol{} & $\frac{3}{5}$ & \barcol{-$\frac{2}{5}$} &
    \barcol{1} & \\
\cline{15-18}
 -$\frac{2}{3}$ & 1 & -$\frac{4}{3}$ & 0 &
    &&&&&&&&&&&& \barcol{} & 1 \\
\hline
\end{tabular}
\end{center}


\spacebetweentables
\tablecaption{10}{Isoscalar factors for {\bf 8} $\mult$ {\bf 3}.}
\begin{center}
\begin{tabular}{rrrr|rrrrrrrrrrr|}
 &&&& $\bar{\bf 6}$ & {\bf 15} & {\bf 3} & $\bar{\bf 6}$ & {\bf 15} &
     {\bf 15} & {\bf 3} & {\bf 15} & $\bar{\bf 6}$ & {\bf 15} & {\bf 15} \\
 \multicolumn{4}{c|}{{\bf 8} $\mult$ {\bf 3}} &
     $\frac{4}{3}$ & $\frac{4}{3}$ & $\frac{1}{3}$ & $\frac{1}{3}$ &
     $\frac{1}{3}$ & $\frac{1}{3}$ & -$\frac{2}{3}$ & -$\frac{2}{3}$ &
     -$\frac{2}{3}$ & -$\frac{2}{3}$ & -$\frac{5}{3}$ \\
 &&&& 0 & 1 & $\frac{1}{2}$ & $\frac{1}{2}$ & $\frac{1}{2}$ & $\frac{3}{2}$ &
    0 & 0 & 1 & 1 & $\frac{1}{2}$ \\
\hline
 1 & $\frac{1}{2}$ & $\frac{1}{3}$ & $\frac{1}{2}$ &
    \barcol{-1} & \barcol{1} & &&&&&&&& \\
\cline{5-9}
 1 & $\frac{1}{2}$ & -$\frac{2}{3}$ & 0 &
    & \barcol{} & $\frac{3}{8}$ & $\frac{1}{4}$ & \barcol{$\frac{3}{8}$} &
    &&&&& \\
 0 & 0 & $\frac{1}{3}$ & $\frac{1}{2}$ &
    & \barcol{} & -$\frac{1}{16}$ & -$\frac{3}{8}$ & \barcol{$\frac{9}{16}$} &
    &&&&& \\
\cline{10-10}
 0 & 1 & $\frac{1}{3}$ & $\frac{1}{2}$ &
    & \barcol{} & $\frac{9}{16}$ & -$\frac{3}{8}$ & \barcol{-$\frac{1}{16}$} &
    \barcol{1}& &&&& \\
\cline{7-12}
 0 & 0 & -$\frac{2}{3}$ & 0 &
    &&&&& \barcol{} & $\frac{1}{4}$ & \barcol{$\frac{3}{4}$} & && \\
\cline{13-14}
 -1 & $\frac{1}{2}$ & $\frac{1}{3}$ & $\frac{1}{2}$ &
    &&&&& \barcol{} & $\frac{3}{4}$ & \barcol{-$\frac{1}{4}$} &
    -$\frac{1}{2}$ & \barcol{$\frac{1}{2}$} & \\
\cline{11-12}
 0 & 1 & -$\frac{2}{3}$ & 0 &
    &&&&&&& \barcol{} & $\frac{1}{2}$ & \barcol{$\frac{1}{2}$} & \\
\cline{13-15}
 -1 & $\frac{1}{2}$ & -$\frac{2}{3}$ & 0 &
    &&&&&&&&& \barcol{} & 1 \\
\hline
\end{tabular}
\end{center}


\newpage
\tablecaption{11}{Isoscalar factors for {\bf 8} $\mult$ {\bf 6}.}
\begin{center}
\begin{tabular}{rrrr|rrrrrrrr|}
 &&&& $\bar{\bf 15}$ & $\bar{\bf 24}$ & $\bar{\bf 3}$ & $\bar{\bf 15}$ &
    {\bf 6} & $\bar{\bf 15}$ & $\bar{\bf 24}$ & $\bar{\bf 24}$ \\
 \multicolumn{4}{c|}{{\bf 8} $\mult$ {\bf 6}} &
    $\frac{5}{3}$ & $\frac{5}{3}$ & $\frac{2}{3}$ & $\frac{2}{3}$ &
    $\frac{2}{3}$ & $\frac{2}{3}$ & $\frac{2}{3}$ & $\frac{2}{3}$ \\
 &&&& $\frac{1}{2}$ & $\frac{3}{2}$ & 0 & 0 & 1 & 1 & 1 & 2 \\
\hline
 1 & $\frac{1}{2}$ & $\frac{2}{3}$ & 1 &
    \barcol{-1} & \barcol{1} & &&&&& \\
\cline{5-11}
 1 & $\frac{1}{2}$ & -$\frac{1}{3}$ & $\frac{1}{2}$ &
    & \barcol{} & -$\frac{1}{4}$ & \barcol{-$\frac{3}{4}$} &
    $\frac{3}{10}$ & $\frac{1}{6}$ & \barcol{$\frac{8}{15}$} & \\
\cline{12-12}
 0 & 1 & $\frac{2}{3}$ & 1 &
    & \barcol{} & -$\frac{3}{4}$ & \barcol{$\frac{1}{4}$} &
    $\frac{3}{5}$ & -$\frac{1}{3}$ & \barcol{-$\frac{1}{15}$} & 1 \\
\cline{7-8}
\cline{12-12}
 0 & 0 & $\frac{2}{3}$ & 1 &
    &&& \barcol{} & -$\frac{1}{10}$ & -$\frac{1}{2}$ &
    \barcol{$\frac{2}{5}$} & \\
\hline
\end{tabular}

\spacebetweensubtables
\begin{tabular}{rrrr|rrrrrrrrrrr|}
 &&&& $\bar{\bf 3}$ & {\bf 6} & $\bar{\bf 15}$ & $\bar{\bf 24}$ &
    $\bar{\bf 15}$ & $\bar{\bf 24}$ & {\bf 6} & $\bar{\bf 24}$ &
    $\bar{\bf 15}$ & $\bar{\bf 24}$ & $\bar{\bf 24}$ \\
 \multicolumn{4}{c|}{{\bf 8} $\mult$ {\bf 6}} &
    -$\frac{1}{3}$ & -$\frac{1}{3}$ & -$\frac{1}{3}$ & -$\frac{1}{3}$ &
    -$\frac{1}{3}$ & -$\frac{1}{3}$ & -$\frac{4}{3}$ & -$\frac{4}{3}$ &
    -$\frac{4}{3}$ & -$\frac{4}{3}$ & -$\frac{7}{3}$ \\
 &&&& $\frac{1}{2}$ & $\frac{1}{2}$ & $\frac{1}{2}$ & $\frac{1}{2}$ &
    $\frac{3}{2}$ & $\frac{3}{2}$ & 0 & 0 & 1 & 1 & $\frac{1}{2}$ \\
\hline
 1 & $\frac{1}{2}$ & -$\frac{4}{3}$ & 0 &
    $\frac{1}{4}$ & $\frac{3}{10}$ & $\frac{1}{4}$ & \barcol{$\frac{1}{5}$} &
    &&&&&& \\
 0 & 0 & -$\frac{1}{3}$ & $\frac{1}{2}$ &
    -$\frac{3}{16}$ & $\frac{1}{40}$ & -$\frac{3}{16}$ &
    \barcol{$\frac{3}{5}$} & &&&&&& \\
\cline{9-10}
 0 & 1 & -$\frac{1}{3}$ & $\frac{1}{2}$ &
    $\frac{3}{16}$ & $\frac{9}{40}$ & -$\frac{25}{48}$ &
    \barcol{-$\frac{1}{15}$} & $\frac{1}{3}$ & \barcol{$\frac{2}{3}$} & &&&& \\
 -1 & $\frac{1}{2}$ & $\frac{2}{3}$ & 1 &
    -$\frac{3}{8}$ & $\frac{9}{20}$ & $\frac{1}{24}$ &
    \barcol{-$\frac{2}{15}$} & -$\frac{2}{3}$ &
    \barcol{$\frac{1}{3}$} & &&&& \\
\cline{5-12}
 0 & 0 & -$\frac{4}{3}$ & 0 &
    &&&&& \barcol{} & $\frac{2}{5}$ & \barcol{$\frac{3}{5}$} & && \\
\cline{13-14}
 -1 & $\frac{1}{2}$ & -$\frac{1}{3}$ & $\frac{1}{2}$ &
    &&&&& \barcol{} & $\frac{3}{5}$ & \barcol{-$\frac{2}{5}$} &
    -$\frac{1}{3}$ & \barcol{$\frac{2}{3}$} & \\
\cline{11-12}
 0 & 1 & -$\frac{4}{3}$ & 0 &
    &&&&&&& \barcol{} & $\frac{2}{3}$ & \barcol{$\frac{1}{3}$} & \\
\cline{13-15}
 -1 & $\frac{1}{2}$ & -$\frac{4}{3}$ & 0 &
    &&&&&&&&& \barcol{} & 1 \\
\hline
\end{tabular}
\end{center}


\spacebetweentables
\tablecaption{12}{Isoscalar factors for {\bf 10} $\mult$ {\bf 3}.}
\begin{center}
\begin{tabular}{rrrr|rrrrrrrrrrr|}
 &&&& {\bf 15} & {\bf 15$'$} & {\bf 15} & {\bf 15} & {\bf 15$'$} & {\bf 15} &
    {\bf 15} & {\bf 15$'$} & {\bf 15} & {\bf 15$'$} & {\bf 15$'$} \\
 \multicolumn{4}{c|}{{\bf 10} $\mult$ {\bf 3}} &
    $\frac{4}{3}$ & $\frac{4}{3}$ & $\frac{1}{3}$ & $\frac{1}{3}$ &
    $\frac{1}{3}$ & -$\frac{2}{3}$ & -$\frac{2}{3}$ & -$\frac{2}{3}$ &
    -$\frac{5}{3}$ & -$\frac{5}{3}$ & -$\frac{8}{3}$ \\
 &&&& 1 & 2 & $\frac{1}{2}$ & $\frac{3}{2}$ & $\frac{3}{2}$ & 0 & 1 & 1 &
    $\frac{1}{2}$ & $\frac{1}{2}$ & 0 \\
\hline
 1 & $\frac{3}{2}$ & $\frac{1}{3}$ & $\frac{1}{2}$ &
    \barcol{-1} & \barcol{1} & &&&&&&&& \\
\cline{5-9}
 0 & 1 & $\frac{1}{3}$ & $\frac{1}{2}$ &
    & \barcol{} & \barcol{-1} & -$\frac{1}{4}$ & \barcol{$\frac{3}{4}$} &
    &&&&& \\
\cline{7-7}
 1 & $\frac{3}{2}$ & -$\frac{2}{3}$ & 0 &
    && \barcol{} & $\frac{3}{4}$ & \barcol{$\frac{1}{4}$} & &&&&& \\
\cline{8-12}
 -1 & $\frac{1}{2}$ & $\frac{1}{3}$ & $\frac{1}{2}$ &
    &&&& \barcol{} & \barcol{-1} & -$\frac{1}{2}$ & \barcol{$\frac{1}{2}$} &
    && \\
\cline{10-10}
 0 & 1 & -$\frac{2}{3}$ & 0 &
    &&&&& \barcol{} & $\frac{1}{2}$ & \barcol{$\frac{1}{2}$} & && \\
\cline{11-14}
 -1 & $\frac{1}{2}$ & -$\frac{2}{3}$ & 0 &
    &&&&&&& \barcol{} & $\frac{1}{4}$ & \barcol{$\frac{3}{4}$} & \\
 -2 & 0 & $\frac{1}{3}$ & $\frac{1}{2}$ &
    &&&&&&& \barcol{} & -$\frac{3}{4}$ & \barcol{$\frac{1}{4}$} & \\
\cline{13-15}
 -2 & 0 & -$\frac{2}{3}$ & 0 &
    &&&&&&&&& \barcol{} & 1 \\
\hline
\end{tabular}
\end{center}


\spacebetweentables
\tablecaption{13}{Isoscalar factors for {\bf 10} $\mult$ $\bar{\bf 3}$.}
\begin{center}
\begin{tabular}{rrrr|rrrrrrrrrrr|}
 &&&& $\bar{\bf 24}$ & {\bf 6} & $\bar{\bf 24}$ & $\bar{\bf 24}$ & {\bf 6} &
    $\bar{\bf 24}$ & $\bar{\bf 24}$ & {\bf 6} & $\bar{\bf 24}$ &
    $\bar{\bf 24}$ & $\bar{\bf 24}$ \\
 \multicolumn{4}{c|}{{\bf 10} $\mult$ $\bar{\bf 3}$} &
    $\frac{5}{3}$ & $\frac{2}{3}$ & $\frac{2}{3}$ & $\frac{2}{3}$ &
    -$\frac{1}{3}$ & -$\frac{1}{3}$ & -$\frac{1}{3}$ & -$\frac{4}{3}$ &
    -$\frac{4}{3}$ & -$\frac{4}{3}$ & -$\frac{7}{3}$ \\
 &&&& $\frac{3}{2}$ & 1 & 1 & 2 & $\frac{1}{2}$ & $\frac{1}{2}$ &
    $\frac{3}{2}$ & 0 & 0 & 1 & $\frac{1}{2}$ \\
\hline
 1 & $\frac{3}{2}$ & $\frac{2}{3}$ & 0 &
    \barcol{1} & &&&&&&&&& \\
\cline{5-7}
 0 & 1 & $\frac{2}{3}$ & 0 &
    \barcol{} & -$\frac{1}{5}$ & \barcol{$\frac{4}{5}$} & &&&&&&& \\
\cline{8-8}
 1 & $\frac{3}{2}$ & -$\frac{1}{3}$ & $\frac{1}{2}$ &
    \barcol{} & $\frac{4}{5}$ & \barcol{$\frac{1}{5}$} & \barcol{1} &
    &&&&&& \\
\cline{6-10}
 -1 & $\frac{1}{2}$ & $\frac{2}{3}$ & 0 &
    &&& \barcol{} & -$\frac{2}{5}$ & \barcol{$\frac{3}{5}$} & &&&& \\
\cline{11-11}
 0 & 1 & -$\frac{1}{3}$ & $\frac{1}{2}$ &
    &&& \barcol{} & $\frac{3}{5}$ & \barcol{$\frac{2}{5}$} & \barcol{1} &
    &&& \\
\cline{9-13}
 -2 & 0 & $\frac{2}{3}$ & 0 &
    &&&&&& \barcol{} & -$\frac{3}{5}$ & \barcol{$\frac{2}{5}$} & & \\
\cline{14-14}
 -1 & $\frac{1}{2}$ & -$\frac{1}{3}$ & $\frac{1}{2}$ &
    &&&&&& \barcol{} & $\frac{2}{5}$ & \barcol{$\frac{3}{5}$} & \barcol{1} &\\
\cline{12-15}
 -2 & 0 & -$\frac{1}{3}$ & $\frac{1}{2}$ &
    &&&&&&&&& \barcol{} & 1 \\
\hline
\end{tabular}
\end{center}


\spacebetweentables
\tablecaption{14}{Isoscalar factors for {\bf 10} $\mult$ {\bf 6}.}
\begin{center}
\begin{tabular}{rrrr|rrrrrrrr|}
 &&&& $\bar{\bf 15}$ & $\bar{\bf 24}$ & $\bar{\bf 21}$ & $\bar{\bf 15}$ &
    $\bar{\bf 15}$ & $\bar{\bf 24}$ & $\bar{\bf 24}$ & $\bar{\bf 21}$ \\
 \multicolumn{4}{c|}{{\bf 10} $\mult$ {\bf 6}} &
    $\frac{5}{3}$ & $\frac{5}{3}$ & $\frac{5}{3}$ & $\frac{2}{3}$ &
    $\frac{2}{3}$ & $\frac{2}{3}$ & $\frac{2}{3}$ & $\frac{2}{3}$ \\
 &&&& $\frac{1}{2}$ & $\frac{3}{2}$ & $\frac{5}{2}$ & 0 & 1 & 1 & 2 & 2 \\
\hline
 1 & $\frac{3}{2}$ & $\frac{2}{3}$ & 1 &
    \barcol{1} & \barcol{-1} & \barcol{1} & &&&& \\
\cline{5-12}
 0 & 1 & $\frac{2}{3}$ & 1 &
    && \barcol{} & \barcol{1} & $\frac{1}{3}$ & \barcol{-$\frac{2}{3}$} &
    -$\frac{2}{5}$ & $\frac{3}{5}$ \\
\cline{8-8}
 1 & $\frac{3}{2}$ & -$\frac{1}{3}$ & $\frac{1}{2}$ &
    &&& \barcol{} & -$\frac{2}{3}$ & \barcol{-$\frac{1}{3}$} &
    $\frac{3}{5}$ & $\frac{2}{5}$ \\
\hline
\end{tabular}

\spacebetweensubtables
\begin{tabular}{rrrr|rrrrrrrrrrrr|}
 &&&& $\bar{\bf 15}$ & $\bar{\bf 24}$ & $\bar{\bf 15}$ & $\bar{\bf 24}$ &
    $\bar{\bf 21}$ & $\bar{\bf 24}$ & $\bar{\bf 15}$ & $\bar{\bf 24}$ &
    $\bar{\bf 21}$ & $\bar{\bf 24}$ & $\bar{\bf 21}$ & $\bar{\bf 21}$ \\
 \multicolumn{4}{c|}{{\bf 10} $\mult$ {\bf 6}} &
    -$\frac{1}{3}$ & -$\frac{1}{3}$ & -$\frac{1}{3}$ & -$\frac{1}{3}$ &
    -$\frac{1}{3}$ & -$\frac{4}{3}$ & -$\frac{4}{3}$ & -$\frac{4}{3}$ &
    -$\frac{4}{3}$ & -$\frac{7}{3}$ & -$\frac{7}{3}$ & -$\frac{10}{3}$ \\
 &&&& $\frac{1}{2}$ & $\frac{1}{2}$ & $\frac{3}{2}$ & $\frac{3}{2}$ &
    $\frac{3}{2}$ & 0 & 1 & 1 & 1 & $\frac{1}{2}$ & $\frac{1}{2}$ & 0 \\
\hline
 0 & 1 & -$\frac{1}{3}$ & $\frac{1}{2}$ &
    -$\frac{1}{3}$ & \barcol{-$\frac{2}{3}$} & -$\frac{1}{3}$ &
    $\frac{1}{15}$ & \barcol{$\frac{3}{5}$} & &&&&&& \\
 -1 & $\frac{1}{2}$ & $\frac{2}{3}$ & 1 &
    $\frac{2}{3}$ & \barcol{-$\frac{1}{3}$} & $\frac{1}{6}$ &
    -$\frac{8}{15}$ & \barcol{$\frac{3}{10}$} & &&&&&& \\
\cline{5-6}
 1 & $\frac{3}{2}$ & -$\frac{4}{3}$ & 0 &
    & \barcol{} & $\frac{1}{2}$ & $\frac{2}{5}$ & \barcol{$\frac{1}{10}$} &
    &&&&&& \\
\cline{7-13}
 -1 & $\frac{1}{2}$ & -$\frac{1}{3}$ & $\frac{1}{2}$ &
    &&&& \barcol{} & \barcol{-1} & -$\frac{1}{3}$ & -$\frac{1}{15}$ &
    \barcol{$\frac{3}{5}$} & && \\
\cline{10-10}
 0 & 1 & -$\frac{4}{3}$ & 0 &
    &&&&& \barcol{} & $\frac{1}{6}$ & $\frac{8}{15}$ &
    \barcol{$\frac{3}{10}$} & && \\
 -2 & 0 & $\frac{2}{3}$ & 1 &
    &&&&& \barcol{} & $\frac{1}{2}$ & -$\frac{2}{5}$ &
    \barcol{$\frac{1}{10}$} & && \\
\cline{11-15}
 -1 & $\frac{1}{2}$ & -$\frac{4}{3}$ & 0 &
    &&&&&&&& \barcol{} & $\frac{2}{5}$ & \barcol{$\frac{3}{5}$} & \\
 -2 & 0 & -$\frac{1}{3}$ & $\frac{1}{2}$ &
    &&&&&&&& \barcol{} & -$\frac{3}{5}$ & \barcol{$\frac{2}{5}$} & \\
\cline{14-16}
 -2 & 0 & -$\frac{4}{3}$ & 0 &
    &&&&&&&&&& \barcol{} & 1 \\
\hline
\end{tabular}
\end{center}


\spacebetweentables
\tablecaption{15}{Isoscalar factors for {\bf 10} $\mult$ $\bar{\bf 6}$.}
\begin{center}
\begin{tabular}{rrrr|rrrrrrrrrr|}
 &&&& {\bf 42} & {\bf 15} & {\bf 42} & {\bf 42} & {\bf 3} & {\bf 15} &
    {\bf 42} & {\bf 15} & {\bf 42} & {\bf 42} \\
 \multicolumn{4}{c|}{{\bf 10} $\mult$ $\bar{\bf 6}$.} &
    $\frac{7}{3}$ & $\frac{4}{3}$ & $\frac{4}{3}$ & $\frac{4}{3}$ &
    $\frac{1}{3}$ & $\frac{1}{3}$ & $\frac{1}{3}$ & $\frac{1}{3}$ &
    $\frac{1}{3}$ & $\frac{1}{3}$ \\
 &&&& $\frac{3}{2}$ & 1 & 1 & 2 & $\frac{1}{2}$ & $\frac{1}{2}$ &
    $\frac{1}{2}$ & $\frac{3}{2}$ & $\frac{3}{2}$ & $\frac{5}{2}$ \\
\hline
 1 & $\frac{3}{2}$ & $\frac{4}{3}$ & 0 &
    \barcol{1} & &&&&&&&& \\
\cline{5-7}
 0 & 1 & $\frac{4}{3}$ & 0 &
    \barcol{} & -$\frac{1}{3}$ & \barcol{$\frac{2}{3}$} & &&&&&& \\
\cline{8-8}
 1 & $\frac{3}{2}$ & $\frac{1}{3}$ & $\frac{1}{2}$ &
    \barcol{} & $\frac{2}{3}$ & \barcol{$\frac{1}{3}$} & \barcol{1} & &&&&& \\
\cline{6-11}
 -1 & $\frac{1}{2}$ & $\frac{4}{3}$ & 0 &
    &&& \barcol{} & $\frac{1}{10}$ & -$\frac{1}{2}$ & \barcol{$\frac{2}{5}$} &
    && \\
\cline{12-13}
 0 & 1 & $\frac{1}{3}$ & $\frac{1}{2}$ &
    &&& \barcol{} & -$\frac{3}{10}$ & $\frac{1}{6}$ & \barcol{$\frac{8}{15}$} &
    -$\frac{1}{6}$ & \barcol{$\frac{5}{6}$} & \\
\cline{14-14}
 1 & $\frac{3}{2}$ & -$\frac{2}{3}$ & 1 &
    &&& \barcol{} & $\frac{3}{5}$ & $\frac{1}{3}$ & \barcol{$\frac{1}{15}$} &
    $\frac{5}{6}$ & \barcol{$\frac{1}{6}$} & 1 \\
\hline
\end{tabular}

\spacebetweensubtables
\begin{tabular}{rrrr|rrrrrrrrrr|}
 &&&& {\bf 3} & {\bf 15} & {\bf 42} & {\bf 15} & {\bf 42} & {\bf 42} &
    {\bf 15} & {\bf 42} & {\bf 42} & {\bf 42} \\
 \multicolumn{4}{c|}{{\bf 10} $\mult$ $\bar{\bf 6}$} &
    -$\frac{2}{3}$ & -$\frac{2}{3}$ & -$\frac{2}{3}$ & -$\frac{2}{3}$ &
    -$\frac{2}{3}$ & -$\frac{2}{3}$ & -$\frac{5}{3}$ & -$\frac{5}{3}$ &
    -$\frac{5}{3}$ & -$\frac{8}{3}$ \\
 &&&& 0 & 0 & 0 & 1 & 1 & 2 & $\frac{1}{2}$ & $\frac{1}{2}$ &
     $\frac{3}{2}$ & 1 \\
\hline
 -2 & 0 & $\frac{4}{3}$ & 0 &
    $\frac{3}{10}$ & -$\frac{1}{2}$ & \barcol{$\frac{1}{5}$} & &&&&&& \\
\cline{8-9}
 -1 & $\frac{1}{2}$ & $\frac{1}{3}$ & $\frac{1}{2}$ &
    -$\frac{2}{5}$ & 0 & \barcol{$\frac{3}{5}$} &
    -$\frac{1}{3}$ & \barcol{$\frac{2}{3}$} & &&&& \\
\cline{10-10}
 0 & 1 & -$\frac{2}{3}$ & 1 &
    $\frac{3}{10}$ & $\frac{1}{2}$ & \barcol{$\frac{1}{5}$} &
    $\frac{2}{3}$ & \barcol{$\frac{1}{3}$} & \barcol{1} & &&& \\
\cline{5-12}
 -2 & 0 & $\frac{1}{3}$ & $\frac{1}{2}$ &
    &&&&& \barcol{} & -$\frac{1}{2}$ & \barcol{$\frac{1}{2}$} & & \\
\cline{13-13}
 -1 & $\frac{1}{2}$ & -$\frac{2}{3}$ & 1 &
    &&&&& \barcol{} & $\frac{1}{2}$ & \barcol{$\frac{1}{2}$} & \barcol{1} & \\
\cline{11-14}
 -2 & 0 & -$\frac{2}{3}$ & 1 &
    &&&&&&&& \barcol{} & 1 \\
\hline
\end{tabular}
\end{center}

\newpage


\spacebetweentables
\tablecaption{16}{Isoscalar factors for {\bf 15} $\mult$ {\bf 3}.}
\begin{center}
\begin{tabular}{rrrr|rrrrrrrrrrrrrrrrr|}
 &&&& $\bar{\bf 15}$ & $\bar{\bf 24}$ & $\bar{\bf 15}$ & {\bf 6} &
    $\bar{\bf 15}$ & $\bar{\bf 24}$ & $\bar{\bf 24}$ &
    {\bf 6} & $\bar{\bf 15}$ & $\bar{\bf 24}$ & $\bar{\bf 15}$ &
    $\bar{\bf 24}$ & {\bf 6} & $\bar{\bf 24}$ & $\bar{\bf 15}$ &
    $\bar{\bf 24}$ & $\bar{\bf 24}$ \\
 \multicolumn{4}{c|}{{\bf 15} $\mult$ {\bf 3}} &
    $\frac{5}{3}$ & $\frac{5}{3}$ & $\frac{2}{3}$ & $\frac{2}{3}$ &
    $\frac{2}{3}$ & $\frac{2}{3}$ & $\frac{2}{3}$ &
    -$\frac{1}{3}$ & -$\frac{1}{3}$ & -$\frac{1}{3}$ & -$\frac{1}{3}$ &
    -$\frac{1}{3}$ & -$\frac{4}{3}$ & -$\frac{4}{3}$ & -$\frac{4}{3}$ &
    -$\frac{4}{3}$ & -$\frac{7}{3}$ \\
 &&&& $\frac{1}{2}$ & $\frac{3}{2}$ & 0 & 1 & 1 & 1 & 2 &
    $\frac{1}{2}$ & $\frac{1}{2}$ & $\frac{1}{2}$ & $\frac{3}{2}$ &
    $\frac{3}{2}$ & 0 & 0 & 1 & 1 & $\frac{1}{2}$ \\
\hline
 $\frac{4}{3}$ & 1 & $\frac{1}{3}$ & $\frac{1}{2}$ &
    \barcol{-1} & \barcol{1} & &&&&&&&&&&&&&& \\
\cline{5-10}
 $\frac{1}{3}$ & $\frac{1}{2}$ & $\frac{1}{3}$ & $\frac{1}{2}$ &
    & \barcol{} & \barcol{-1} & -$\frac{1}{15}$ & -$\frac{2}{9}$ &
    \barcol{$\frac{32}{45}$} &&&&&&&&&&& \\
\cline{7-7}
 $\frac{4}{3}$ & 1 & -$\frac{2}{3}$ & 0 &
    && \barcol{} & $\frac{2}{5}$ & $\frac{1}{3}$ & \barcol{$\frac{4}{15}$} &
    &&&&&&&&&& \\
\cline{11-11}
 $\frac{1}{3}$ & $\frac{3}{2}$ & $\frac{1}{3}$ & $\frac{1}{2}$ &
    && \barcol{} & $\frac{8}{15}$ & -$\frac{4}{9}$ &
    \barcol{-$\frac{1}{45}$} & \barcol{1} & &&&&&&&&& \\
\cline{8-14}
 $\frac{1}{3}$ & $\frac{1}{2}$ & -$\frac{2}{3}$ & 0 &
    &&&&&& \barcol{} &
    $\frac{3}{10}$ & $\frac{1}{6}$ & \barcol{$\frac{8}{15}$} & &&&&&& \\
\hline
\end{tabular}
\end{center}


\spacebetweentables
\tablecaption{17}{Isoscalar factors for {\bf 15} $\mult$ $\bar{\bf 3}$.}
\begin{center}
\begin{tabular}{rrrr|rrrrrrrrrrrrrrrrr|}
 & & & & {\bf 27} & {\bf 8} & {\bf 27} & {\bf 10} & {\bf 27} &
    {\bf 8} & {\bf 27} & {\bf 8} & {\bf 10} & {\bf 27} & {\bf 27} &
    {\bf 8} & {\bf 10} & {\bf 27} & {\bf 27} & {\bf 10} & {\bf 27} \\
\multicolumn{4}{c|}{{\bf 15} $\mult$ $\bar{\bf 3}$} &
       2 & 1 & 1 & 1 & 1 &
       0 & 0 & 0 & 0 & 0 & 0 &
       -1 & -1 & -1 & -1 & -2 & -2 \\
 & & & & 1 & $\frac{1}{2}$ & $\frac{1}{2}$ & $\frac{3}{2}$ & $\frac{3}{2}$ &
    0 & 0 & 1 & 1 & 1 & 2 &
    $\frac{1}{2}$ & $\frac{1}{2}$ & $\frac{1}{2}$ &
    $\frac{3}{2}$ & 0 & 1 \\
\hline
 $\frac{4}{3}$ & 1 & $\frac{2}{3}$ & 0 & \multicolumn{1}{r|}{1} & & & &
    &&&&&&&&&&&& \\
\cline{5-7}
 $\frac{1}{3}$ & $\frac{1}{2}$ & $\frac{2}{3}$ & 0 & \multicolumn{1}{r|}{} &
                   -$\frac{1}{5}$ & \multicolumn{1}{r|}{$\frac{4}{5}$} & &
    &&&&&&&&&&&& \\
\cline{8-9}
 $\frac{4}{3}$ & 1 & -$\frac{1}{3}$ & $\frac{1}{2}$ & \multicolumn{1}{r|}{} &
                   $\frac{4}{5}$ & \multicolumn{1}{r|}{$\frac{1}{5}$} &
                   -$\frac{1}{2}$ & \multicolumn{1}{r|}{$\frac{1}{2}$}
    &&&&&&&&&&&& \\
\cline{6-7}
 $\frac{1}{3}$ & $\frac{3}{2}$ & $\frac{2}{3}$ & 0 &
                 & & \multicolumn{1}{r|}{} &
        $\frac{1}{2}$ & \multicolumn{1}{r|}{$\frac{1}{2}$}
    &&&&&&&&&&&& \\
\cline{8-11}
 -$\frac{2}{3}$ & 0 & $\frac{2}{3}$ & 0 &
       &&&& \barcol{} &
          -$\frac{2}{5}$ & \multicolumn{1}{r|}{$\frac{3}{5}$} & &&&&&&&&& \\
\hline
\end{tabular}
\end{center}


\spacebetweentables
\tablecaption{18}{Isoscalar factors for {\bf 15} $\mult$ {\bf 6}.}
\begin{center}
\begin{tabular}{rrrr|rrrrrrrrrr|}
 &&&& $\bar{\bf 10}$ & {\bf 27} & {\bf 35} & {\bf 8} & $\bar{\bf 10}$ &
    {\bf 27} & {\bf 10} & {\bf 27} & {\bf 35} & {\bf 35} \\
 \multicolumn{4}{c|}{{\bf 15} $\mult$ {\bf 6}} &
    2 & 2 & 2 & 1 & 1 & 1 & 1 & 1 & 1 & 1 \\
 &&&& 0 & 1 & 2 & $\frac{1}{2}$ & $\frac{1}{2}$ & $\frac{1}{2}$ &
    $\frac{3}{2}$ & $\frac{3}{2}$ & $\frac{3}{2}$ & $\frac{5}{2}$ \\
\hline
 $\frac{4}{3}$ & 1 & $\frac{2}{3}$ & 1 &
    \barcol{1} & \barcol{-1} & \barcol{1} & &&&&&& \\
\cline{5-13}
 $\frac{4}{3}$ & 1 & -$\frac{1}{3}$ & $\frac{1}{2}$ &
    && \barcol{} & -$\frac{4}{15}$ & -$\frac{1}{3}$ & \barcol{-$\frac{2}{5}$} &
    $\frac{1}{3}$ & $\frac{1}{4}$ & \barcol{$\frac{5}{12}$} & \\
 $\frac{1}{3}$ & $\frac{1}{2}$ & $\frac{2}{3}$ & 1 &
    && \barcol{} & $\frac{1}{45}$ & $\frac{4}{9}$ & \barcol{-$\frac{8}{15}$} &
    -$\frac{1}{9}$ & -$\frac{1}{3}$ & \barcol{$\frac{5}{9}$} & \\
\cline{14-14}
 $\frac{1}{3}$ & $\frac{3}{2}$ & $\frac{2}{3}$ & 1 &
    && \barcol{} & -$\frac{32}{45}$ & $\frac{2}{9}$ & \barcol{$\frac{1}{15}$} &
    $\frac{5}{9}$ & -$\frac{5}{12}$ & \barcol{-$\frac{1}{36}$} & 1 \\
\hline
\end{tabular}

\spacebetweensubtables
\begin{tabular}{rrrr|rrrrrrrrr|}
 &&&& {\bf 8} & {\bf 27} & {\bf 8} & {\bf 10} & $\bar{\bf 10}$ &
    {\bf 27} & {\bf 35} & {\bf 27} & {\bf 35} \\
 \multicolumn{4}{c|}{{\bf 15} $\mult$ {\bf 6}} &
    0 & 0 & 0 & 0 & 0 & 0 & 0 & 0 & 0 \\
 &&&& 0 & 0 & 1 & 1 & 1 & 1 & 1 & 2 & 2 \\
\hline
 $\frac{1}{3}$ & $\frac{1}{2}$ & -$\frac{1}{3}$ & $\frac{1}{2}$ &
    -$\frac{1}{5}$ & \barcol{-$\frac{4}{5}$} &
    -$\frac{5}{27}$ & $\frac{2}{27}$ &
    -$\frac{4}{27}$ & 0 & \barcol{$\frac{16}{27}$} & & \\
\cline{12-13}
 -$\frac{2}{3}$ & 1 & $\frac{2}{3}$ & 1 &
    -$\frac{4}{5}$ & \barcol{$\frac{1}{5}$} & -$\frac{8}{45}$ & $\frac{4}{9}$ &
    $\frac{2}{9}$ & -$\frac{1}{10}$ & \barcol{-$\frac{1}{18}$} &
    -$\frac{1}{2}$ & $\frac{1}{2}$ \\
\cline{5-6}
 $\frac{1}{3}$ & $\frac{3}{2}$ & -$\frac{1}{3}$ & $\frac{1}{2}$ &
    & \barcol{} & $\frac{32}{135}$ & $\frac{4}{27}$ & -$\frac{8}{27}$ &
    -$\frac{3}{10}$ & \barcol{-$\frac{1}{54}$} &
    $\frac{1}{2}$ & $\frac{1}{2}$ \\
\cline{12-13}
 $\frac{4}{3}$ & 1 & -$\frac{4}{3}$ & 0 &
    & \barcol{} & $\frac{16}{45}$ & $\frac{2}{9}$ & $\frac{1}{9}$ &
    $\frac{1}{5}$ & \barcol{$\frac{1}{9}$} & & \\
 -$\frac{2}{3}$ & 0 & $\frac{2}{3}$ & 1 &
    & \barcol{} & $\frac{2}{45}$ & -$\frac{1}{9}$ & $\frac{2}{9}$ &
    -$\frac{2}{5}$ & \barcol{$\frac{2}{9}$} & & \\
\hline
\end{tabular}

\spacebetweensubtables
\begin{tabular}{rrrr|rrrrrrrrrrrr|}
 &&&& {\bf 8} & {\bf 10} & {\bf 27} & {\bf 35} & $\bar{\bf 10}$ & {\bf 27} &
    {\bf 35} & {\bf 10} & {\bf 35} & {\bf 27} & {\bf 35} & {\bf 35} \\
 \multicolumn{4}{c|}{{\bf 15} $\mult$ {\bf 6}} &
    -1 & -1 & -1 & -1 & -1 & -1 & -1 & -2 & -2 & -2 & -2 & -3 \\
 &&&& $\frac{1}{2}$ & $\frac{1}{2}$ & $\frac{1}{2}$ & $\frac{1}{2}$ &
    $\frac{3}{2}$ & $\frac{3}{2}$ & $\frac{3}{2}$ &
    0 & 0 & 1 & 1 & $\frac{1}{2}$ \\
\hline
 $\frac{1}{3}$ & $\frac{1}{2}$ & -$\frac{4}{3}$ & 0 &
    $\frac{2}{15}$ & $\frac{1}{3}$ & $\frac{1}{5}$ & \barcol{$\frac{1}{3}$} &
    &&&&&&& \\
 -$\frac{2}{3}$ & 0 & -$\frac{1}{3}$ & $\frac{1}{2}$ &
    -$\frac{1}{5}$ & 0 & -$\frac{3}{10}$ & \barcol{$\frac{1}{2}$} &
    &&&&&&& \\
\cline{9-11}
 -$\frac{2}{3}$ & 1 & -$\frac{1}{3}$ & $\frac{1}{2}$ &
    $\frac{2}{15}$ & $\frac{1}{3}$ & -$\frac{9}{20}$ &
    \barcol{-$\frac{1}{12}$} & -$\frac{1}{3}$ & 0 & \barcol{$\frac{2}{3}$} &
    &&&& \\
 -$\frac{5}{3}$ & $\frac{1}{2}$ & $\frac{2}{3}$ & 1 &
    -$\frac{8}{15}$ & $\frac{1}{3}$ & $\frac{1}{20}$ &
    \barcol{-$\frac{1}{12}$} & $\frac{1}{3}$ & -$\frac{1}{2}$ &
    \barcol{$\frac{1}{6}$} & &&&& \\
\cline{5-8}
 $\frac{1}{3}$ & $\frac{3}{2}$ & -$\frac{4}{3}$ & 0 &
    &&& \barcol{} & $\frac{1}{3}$ & $\frac{1}{2}$ & \barcol{$\frac{1}{6}$} &
    &&&& \\
\cline{9-13}
 -$\frac{2}{3}$ & 0 & -$\frac{4}{3}$ & 0 &
    &&&&&& \barcol{} & $\frac{1}{3}$ & \barcol{$\frac{2}{3}$} & && \\
\cline{14-15}
 -$\frac{5}{3}$ & $\frac{1}{2}$ & -$\frac{1}{3}$ & $\frac{1}{2}$ &
    &&&&&& \barcol{} & $\frac{2}{3}$ & \barcol{-$\frac{1}{3}$} &
    -$\frac{1}{2}$ & \barcol{$\frac{1}{2}$} & \\
\cline{12-13}
 -$\frac{2}{3}$ & 1 & -$\frac{4}{3}$ & 0 &
    &&&&&&&& \barcol{} & $\frac{1}{2}$ & \barcol{$\frac{1}{2}$} & \\
\cline{14-16}
 -$\frac{5}{3}$ & $\frac{1}{2}$ & -$\frac{4}{3}$ & 0 &
    &&&&&&&&&& \barcol{} & 1 \\
\hline
\end{tabular}
\end{center}


\newpage

\spacebetweentables
\tablecaption{19}{Isoscalar factors for {\bf 15} $\mult$ $\bar{\bf 6}$.}
\begin{center}
\begin{tabular}{rrrr|rrrrrrrrrrrrrr|}
 &&&& $\bar{\bf 42}$ & $\bar{\bf 15}$ & $\bar{\bf 42}$ & $\bar{\bf 24}$ &
    $\bar{\bf 42}$ & $\bar{\bf 3}$ & $\bar{\bf 15}$ & $\bar{\bf 42}$ &
    {\bf 6} & $\bar{\bf 15}$ & $\bar{\bf 24}$ & $\bar{\bf 42}$ &
    $\bar{\bf 24}$ & $\bar{\bf 42}$ \\
 \multicolumn{4}{c|}{{\bf 15} $\mult$ $\bar{\bf 6}$} &
    $\frac{8}{3}$ & $\frac{5}{3}$ & $\frac{5}{3}$ & $\frac{5}{3}$ &
    $\frac{5}{3}$ & $\frac{2}{3}$ & $\frac{2}{3}$ & $\frac{2}{3}$ &
    $\frac{2}{3}$ & $\frac{2}{3}$ & $\frac{2}{3}$ & $\frac{2}{3}$ &
    $\frac{2}{3}$ & $\frac{2}{3}$ \\
 &&&& 1 & $\frac{1}{2}$ & $\frac{1}{2}$ & $\frac{3}{2}$ & $\frac{3}{2}$ &
    0 & 0 & 0 & 1 & 1 & 1 & 1 & 2 & 2 \\
\hline
 $\frac{4}{3}$ & 1 & $\frac{4}{3}$ & 0 &
    \barcol{1} & &&&&&&&&&&&& \\
\cline{5-7}
 $\frac{1}{3}$ & $\frac{1}{2}$ & $\frac{4}{3}$ & 0 &
    \barcol{} & -$\frac{1}{3}$ & \barcol{$\frac{2}{3}$} & &&&&&&&&&& \\
\cline{8-9}
 $\frac{4}{3}$ & 1 & $\frac{1}{3}$ & $\frac{1}{2}$ &
    \barcol{} & $\frac{2}{3}$ & \barcol{$\frac{1}{3}$} &
    -$\frac{1}{3}$ & \barcol{$\frac{2}{3}$} & &&&&&&&& \\
\cline{6-7}
 $\frac{1}{3}$ & $\frac{3}{2}$ & $\frac{4}{3}$ & 0 &
    && \barcol{} & $\frac{2}{3}$ & \barcol{$\frac{1}{3}$} & &&&&&&&& \\
\cline{8-12}
 -$\frac{2}{3}$ & 0 & $\frac{4}{3}$ & 0 &
    &&&& \barcol{} & $\frac{1}{10}$ & -$\frac{1}{2}$ & \barcol{$\frac{2}{5}$} &
    &&&&& \\
\hline
\end{tabular}

\spacebetweensubtables
\begin{tabular}{rrrr|rrrrrrrrrrrrrrrrr|}
 &&&& $\bar{\bf 3}$ & {\bf 6} & $\bar{\bf 15}$ & $\bar{\bf 24}$ &
    $\bar{\bf 42}$ & $\bar{\bf 15}$ & $\bar{\bf 24}$ & $\bar{\bf 42}$ &
    $\bar{\bf 42}$ & {\bf 6} & $\bar{\bf 24}$ & $\bar{\bf 15}$ &
    $\bar{\bf 24}$ & $\bar{\bf 42}$ & $\bar{\bf 42}$ & $\bar{\bf 24}$ &
    $\bar{\bf 42}$ \\
 \multicolumn{4}{c|}{{\bf 15} $\mult$ $\bar{\bf 6}$} &
    -$\frac{1}{3}$ & -$\frac{1}{3}$ & -$\frac{1}{3}$ & -$\frac{1}{3}$ &
    -$\frac{1}{3}$ & -$\frac{1}{3}$ & -$\frac{1}{3}$ & -$\frac{1}{3}$ &
    -$\frac{1}{3}$ & -$\frac{4}{3}$ & -$\frac{4}{3}$ & -$\frac{4}{3}$ &
    -$\frac{4}{3}$ & -$\frac{4}{3}$ & -$\frac{4}{3}$ &
    -$\frac{7}{3}$ & -$\frac{7}{3}$ \\
 &&&& $\frac{1}{2}$ & $\frac{1}{2}$ & $\frac{1}{2}$ & $\frac{1}{2}$ &
    $\frac{1}{2}$ & $\frac{3}{2}$ & $\frac{3}{2}$ & $\frac{3}{2}$ &
    $\frac{5}{2}$ & 0 & 0 & 1 & 1 & 1 & 2 & $\frac{1}{2}$ & $\frac{3}{2}$ \\
\hline
 -$\frac{5}{3}$ & $\frac{1}{2}$ & $\frac{4}{3}$ & 0 &
    $\frac{1}{5}$ & -$\frac{1}{5}$ & -$\frac{1}{3}$ & $\frac{2}{15}$ &
    \barcol{$\frac{2}{15}$} & &&&&&&&&&&& \\
 -$\frac{2}{3}$ & 0 & $\frac{1}{3}$ & $\frac{1}{2}$ &
    -$\frac{1}{20}$ & $\frac{1}{5}$ & -$\frac{1}{12}$ & -$\frac{2}{15}$ &
    \barcol{$\frac{8}{15}$} & &&&&&&&&&&& \\
\cline{10-12}
 -$\frac{2}{3}$ & 1 & $\frac{1}{3}$ & $\frac{1}{2}$ &
    -$\frac{3}{10}$ & 0 & $\frac{1}{18}$ & $\frac{5}{9}$ &
    \barcol{$\frac{4}{45}$} & -$\frac{2}{9}$ & $\frac{2}{9}$ &
    \barcol{$\frac{5}{9}$} & &&&&&&&& \\
 $\frac{1}{3}$ & $\frac{1}{2}$ & -$\frac{2}{3}$ & 1 &
    $\frac{1}{20}$ & -$\frac{1}{5}$ & $\frac{49}{108}$ & -$\frac{8}{135}$ &
    \barcol{$\frac{32}{135}$} & $\frac{1}{27}$ & -$\frac{16}{27}$ &
    \barcol{$\frac{10}{27}$} & &&&&&&&& \\
\cline{13-13}
 $\frac{1}{3}$ & $\frac{3}{2}$ & -$\frac{2}{3}$ & 1 &
    $\frac{2}{5}$ & $\frac{2}{5}$ & $\frac{2}{27}$ & $\frac{16}{135}$ &
    \barcol{$\frac{1}{135}$} & $\frac{20}{27}$ & $\frac{5}{27}$ &
    \barcol{$\frac{2}{27}$} & \barcol{1} & &&&&&&& \\
\hline
\end{tabular}
\end{center}


\newpage

\spacebetweentables
\tablecaption{20}{Isoscalar factors for {\bf 15} $\mult$ {\bf 8}.}
\begin{center}
\begin{tabular}{rrrr|rrrrrrrrrr|}
 &&&& {\bf 24} & {\bf 42} & $\bar{\bf 6}$ & {\bf 24} & {\bf 15}$_1$ &
    {\bf 15}$_2$ & {\bf 24} & {\bf 42} & {\bf 15$'$} & {\bf 42} \\
 \multicolumn{4}{c|}{{\bf 15} $\mult$ {\bf 8}} &
    $\frac{7}{3}$ & $\frac{7}{3}$ &
    $\frac{4}{3}$ & $\frac{4}{3}$ & $\frac{4}{3}$ & $\frac{4}{3}$ &
    $\frac{4}{3}$ & $\frac{4}{3}$ & $\frac{4}{3}$ & $\frac{4}{3}$ \\
 &&&& $\frac{1}{2}$ & $\frac{3}{2}$ & 0 & 0 & 1 & 1 & 1 & 1 & 2 & 2 \\
\hline
 $\frac{4}{3}$ & 1 & 1 & $\frac{1}{2}$ &
    \barcol{-1} & \barcol{1} & &&&&&&& \\
\cline{5-12}
 $\frac{1}{3}$ & $\frac{1}{2}$ & 1 & $\frac{1}{2}$ &
    & \barcol{} & $\frac{1}{5}$ & \barcol{-$\frac{4}{5}$} &
    -$\frac{1}{61}$ & -$\frac{400}{1647}$ & -$\frac{4}{27}$ &
    \barcol{$\frac{16}{27}$} & & \\
\cline{13-14}
 $\frac{4}{3}$ & 1 & 0 & 1 &
    & \barcol{} & -$\frac{4}{5}$ & \barcol{-$\frac{1}{5}$} &
    $\frac{24}{61}$ & $\frac{361}{1098}$ & -$\frac{2}{9}$ &
    \barcol{$\frac{1}{18}$} & -$\frac{1}{2}$ & $\frac{1}{2}$ \\
\cline{7-8}
 $\frac{1}{3}$ & $\frac{3}{2}$ & 1 & $\frac{1}{2}$ &
    &&& \barcol{} & -$\frac{32}{61}$ & $\frac{529}{3294}$ &
    -$\frac{8}{27}$ & \barcol{-$\frac{1}{54}$} &
    $\frac{1}{2}$ & $\frac{1}{2}$ \\
\cline{13-14}
 $\frac{4}{3}$ & 1 & 0 & 0 &
    &&& \barcol{} & -$\frac{4}{61}$ & $\frac{49}{183}$ &
    $\frac{1}{3}$ & \barcol{$\frac{1}{3}$} & & \\
\hline
\end{tabular}

\spacebetweensubtables
\begin{tabular}{rrrr|rrrrrrrrrrrr|}
 &&&& {\bf 3} & $\bar{\bf 6}$ & {\bf 15}$_1$ & {\bf 15}$_2$ & {\bf 24} &
    {\bf 42} & {\bf 15}$_1$ & {\bf 15}$_2$ & {\bf 15$'$} & {\bf 24} &
    {\bf 42} & {\bf 42} \\
 \multicolumn{4}{c|}{{\bf 15} $\mult$ {\bf 8}} &
    $\frac{1}{3}$ & $\frac{1}{3}$ & $\frac{1}{3}$ & $\frac{1}{3}$ &
    $\frac{1}{3}$ & $\frac{1}{3}$ & $\frac{1}{3}$ & $\frac{1}{3}$ &
    $\frac{1}{3}$ & $\frac{1}{3}$ & $\frac{1}{3}$ & $\frac{1}{3}$ \\
 &&&& $\frac{1}{2}$ & $\frac{1}{2}$ & $\frac{1}{2}$ & $\frac{1}{2}$ &
    $\frac{1}{2}$ & $\frac{1}{2}$ & $\frac{3}{2}$ & $\frac{3}{2}$ &
    $\frac{3}{2}$ & $\frac{3}{2}$ & $\frac{3}{2}$ & $\frac{5}{2}$ \\
\hline
 $\frac{1}{3}$ & $\frac{1}{2}$ & 0 & 0 &
    -$\frac{9}{80}$ & -$\frac{3}{40}$ & -$\frac{121}{976}$ & $\frac{4}{183}$ &
    $\frac{2}{15}$ & \barcol{$\frac{8}{15}$} & &&&&& \\
 -$\frac{2}{3}$ & 0 & 1 & $\frac{1}{2}$ &
    $\frac{1}{40}$ & $\frac{3}{20}$ & -$\frac{9}{488}$ & -$\frac{150}{549}$ &
    -$\frac{4}{15}$ & \barcol{$\frac{4}{15}$} & &&&&& \\
\cline{11-15}
 $\frac{4}{3}$ & 1 & -1 & $\frac{1}{2}$ &
    $\frac{3}{10}$ & $\frac{1}{5}$ & $\frac{3}{122}$ & $\frac{200}{549}$ &
    $\frac{1}{45}$ & \barcol{$\frac{4}{45}$} &
    -$\frac{24}{61}$ & $\frac{529}{4392}$ & -$\frac{1}{8}$ &
    $\frac{2}{9}$ & \barcol{$\frac{5}{36}$} & \\
 -$\frac{2}{3}$ & 1 & 1 & $\frac{1}{2}$ &
    -$\frac{3}{20}$ & $\frac{1}{10}$ & -$\frac{75}{224}$ & $\frac{121}{549}$ &
    -$\frac{8}{45}$ & \barcol{-$\frac{2}{45}$} &
    -$\frac{12}{61}$ & -$\frac{361}{2196}$ & $\frac{1}{4}$ &
    -$\frac{1}{9}$ & \barcol{$\frac{5}{18}$} & \\
 $\frac{1}{3}$ & $\frac{1}{2}$ & 0 & 1 &
    -$\frac{1}{80}$ & -$\frac{5}{24}$ & $\frac{225}{976}$ &
    $\frac{196}{1647}$ & -$\frac{10}{27}$ & \barcol{$\frac{8}{135}$} &
    $\frac{9}{61}$ & -$\frac{1}{1647}$ & -$\frac{1}{3}$ &
    -$\frac{4}{27}$ & \barcol{$\frac{10}{27}$} & \\
\hline
\end{tabular}

\spacebetweensubtables
\begin{tabular}{rrrr|rrrrrrrrrrrr|}
 &&&& {\bf 3} & {\bf 15}$_1$ & {\bf 15}$_2$ & {\bf 42} & $\bar{\bf 6}$ &
    {\bf 15}$_1$ & {\bf 15}$_2$ & {\bf 15$'$} & {\bf 24} & {\bf 42} &
    {\bf 24} & {\bf 42} \\
 \multicolumn{4}{c|}{{\bf 15} $\mult$ {\bf 8}} &
    -$\frac{2}{3}$ & -$\frac{2}{3}$ & -$\frac{2}{3}$ & -$\frac{2}{3}$ &
    -$\frac{2}{3}$ & -$\frac{2}{3}$ & -$\frac{2}{3}$ & -$\frac{2}{3}$ &
    -$\frac{2}{3}$ & -$\frac{2}{3}$ & -$\frac{2}{3}$ & -$\frac{2}{3}$ \\
 &&&& 0 & 0 & 0 & 0 & 1 & 1 & 1 & 1 & 1 & 1 & 2 & 2 \\
\hline
 -$\frac{2}{3}$ & 0 & 0 & 0 &
    -$\frac{3}{20}$ & -$\frac{49}{244}$ & -$\frac{3}{61}$ &
    \barcol{$\frac{3}{5}$} & &&&&&&& \\
\cline{9-14}
 -$\frac{5}{3}$ & $\frac{1}{2}$ & 1 & $\frac{1}{2}$ &
    -$\frac{2}{5}$ & -$\frac{6}{61}$ & $\frac{49}{122}$ &
    \barcol{-$\frac{1}{10}$} &
    $\frac{4}{15}$ & -$\frac{12}{61}$ & -$\frac{361}{2196}$ & $\frac{1}{12}$ &
    -$\frac{8}{45}$ & \barcol{$\frac{1}{9}$} & & \\
 $\frac{1}{3}$ & $\frac{1}{2}$ & -1 & $\frac{1}{2}$ &
    $\frac{3}{20}$ & $\frac{9}{244}$ & $\frac{100}{183}$ &
    \barcol{$\frac{4}{15}$} &
    $\frac{1}{90}$ & -$\frac{25}{122}$ & $\frac{242}{1647}$ & -$\frac{2}{9}$ &
    $\frac{16}{135}$ & \barcol{$\frac{8}{27}$} & & \\
\cline{15-16}
 -$\frac{2}{3}$ & 1 & 0 & 1 &
    $\frac{3}{10}$ & -$\frac{81}{122}$ & $\frac{1}{366}$ &
    \barcol{-$\frac{1}{30}$} &
    -$\frac{2}{15}$ & -$\frac{6}{61}$ & $\frac{49}{122}$ & $\frac{1}{6}$ &
    -$\frac{1}{5}$ & \barcol{0} &
    -$\frac{1}{3}$ & $\frac{2}{3}$ \\
\cline{5-8}
 $\frac{1}{3}$ & $\frac{3}{2}$ & -1 & $\frac{1}{2}$ &
    &&& \barcol{} & $\frac{16}{45}$ & $\frac{16}{61}$ & $\frac{361}{1647}$ &
    $\frac{1}{9}$ & $\frac{2}{135}$ & \barcol{$\frac{1}{27}$} &
    $\frac{2}{3}$ & $\frac{1}{3}$ \\
\hline
\end{tabular}

\spacebetweensubtables
\begin{tabular}{rrrr|rrrrrrrr|}
 &&&& {\bf 15}$_1$ & {\bf 15}$_2$ & {\bf 15$'$} & {\bf 42} & {\bf 24} &
    {\bf 42} & {\bf 15$'$} & {\bf 42} \\
 \multicolumn{4}{c|}{{\bf 15} $\mult$ {\bf 8}} &
    -$\frac{5}{3}$ & -$\frac{5}{3}$ & -$\frac{5}{3}$ & -$\frac{5}{3}$ &
    -$\frac{5}{3}$ & -$\frac{5}{3}$ & -$\frac{8}{3}$ & -$\frac{8}{3}$ \\
 &&&& $\frac{1}{2}$ & $\frac{1}{2}$ & $\frac{1}{2}$ & $\frac{1}{2}$ &
    $\frac{3}{2}$ & $\frac{3}{2}$ & 0 & 1 \\
\hline
 -$\frac{5}{3}$ & $\frac{1}{2}$ & 0 & 0 &
    -$\frac{4}{61}$ & -$\frac{363}{976}$ & $\frac{3}{16}$ &
    \barcol{$\frac{3}{8}$} & &&& \\
 -$\frac{2}{3}$ & 0 & -1 & $\frac{1}{2}$ &
    -$\frac{3}{61}$ & $\frac{49}{244}$ & -$\frac{1}{4}$ &
    \barcol{$\frac{1}{2}$} & &&& \\
\cline{9-10}
 -$\frac{2}{3}$ & 1 & -1 & $\frac{1}{2}$ &
    $\frac{18}{61}$ & $\frac{361}{1464}$ & $\frac{3}{8}$ &
    \barcol{$\frac{1}{12}$} & $\frac{1}{3}$ & \barcol{$\frac{2}{3}$} & & \\
 -$\frac{5}{3}$ & $\frac{1}{2}$ & 0 & 1 &
    -$\frac{36}{61}$ & $\frac{529}{2928}$ & $\frac{3}{16}$ &
    \barcol{-$\frac{1}{24}$} & -$\frac{2}{3}$ & \barcol{$\frac{1}{3}$} & & \\
\cline{5-12}
 -$\frac{5}{3}$ & $\frac{1}{2}$ & -1 & $\frac{1}{2}$ &
    &&&&& \barcol{} & \barcol{1} & 1 \\
\hline
\end{tabular}
\end{center}


\spacebetweentables
\tablecaption{21}{Isoscalar factors for {\bf 15} $\mult$ {\bf 10}.}
\begin{center}
\begin{tabular}{rrrr|rrrrrrrrrrrr|}
 &&&& {\bf 24} & {\bf 42} & {\bf 48} & $\bar{\bf 6}$ & {\bf 24} & {\bf 15} &
    {\bf 24} & {\bf 42} & {\bf 15$'$} & {\bf 42} & {\bf 48} & {\bf 48} \\
 \multicolumn{4}{c|}{{\bf 15} $\mult$ {\bf 10}} &
    $\frac{7}{3}$ & $\frac{7}{3}$ & $\frac{7}{3}$ & $\frac{4}{3}$ &
    $\frac{4}{3}$ & $\frac{4}{3}$ & $\frac{4}{3}$ & $\frac{4}{3}$ &
    $\frac{4}{3}$ & $\frac{4}{3}$ & $\frac{4}{3}$ & $\frac{4}{3}$ \\
 &&&& $\frac{1}{2}$ & $\frac{3}{2}$ & $\frac{5}{2}$ & 0 & 0 &
    1 & 1 & 1 & 2 & 2 & 2 & 3 \\
\hline
 $\frac{4}{3}$ & 1 & 1 & $\frac{3}{2}$ &
    \barcol{1} & \barcol{-1} & \barcol{1} & &&&&&&&& \\
\cline{5-15}
 $\frac{4}{3}$ & 1 & 0 & 1 &
    && \barcol{} & $\frac{1}{5}$ & \barcol{$\frac{4}{5}$} &
    -$\frac{2}{9}$ & -$\frac{2}{9}$ & \barcol{-$\frac{5}{9}$} &
    $\frac{2}{7}$ & $\frac{1}{5}$ & \barcol{$\frac{18}{35}$} & \\
\cline{16-16}
 $\frac{1}{3}$ & $\frac{3}{2}$ & 1 & $\frac{3}{2}$ &
    && \barcol{} & $\frac{4}{5}$ & \barcol{-$\frac{1}{5}$} &
    -$\frac{20}{27}$ & $\frac{5}{27}$ & \barcol{$\frac{2}{27}$} &
    $\frac{4}{7}$ & -$\frac{2}{5}$ & \barcol{-$\frac{1}{35}$} & 1 \\
\cline{8-9}
\cline{16-16}
 $\frac{1}{3}$ & $\frac{1}{2}$ & 1 & $\frac{3}{2}$ &
    &&&& \barcol{} & $\frac{1}{27}$ & $\frac{16}{27}$ &
    \barcol{-$\frac{10}{27}$} & -$\frac{1}{7}$ & -$\frac{2}{5}$ &
    \barcol{$\frac{16}{35}$} & \\
\hline
\end{tabular}%
\end{center}

\spacebetweentables
\tablecaption{22}{Isoscalar factors for {\bf 15$'$} $\mult$ {\bf 3}.}
\begin{center}
%
\end{center}


\newpage

\spacebetweentables
\tablecaption{23}{Isoscalar factors for {\bf 15$'$} $\mult$ $\bar{\bf 3}$.}
\begin{center}
%
\end{center}


\spacebetweentables
\tablecaption{24}{Isoscalar factors for {\bf 15$'$} $\mult$ {\bf 6}.}
\begin{center}
%
\end{center}


\spacebetweentables
\tablecaption{25}{Isoscalar factors for {\bf 15$'$} $\mult$ $\bar{\bf 6}$.}
\begin{center}
%
\end{center}


\spacebetweentables
\tablecaption{26}{Isoscalar factors for {\bf 15$'$} $\mult$ {\bf 8}.}
\begin{center}
%
\end{center}


\newpage

\spacebetweentables
\tablecaption{27}{Isoscalar factors for {\bf 15$'$} $\mult$ {\bf 10}.}
\begin{center}
%
\end{center}


\newpage

\spacebetweentables
\tablecaption{28}{Isoscalar factors for {\bf 15$'$} $\mult$ {\bf 15$'$}.}
\begin{center}
%
\end{center}

\spacebetweentables
\tablecaption{29}{Isoscalar factors for $\bar{\bf 21}$ $\mult$ {\bf 3}.}
\begin{center}
%
\end{center}


\newpage

\spacebetweentables
\tablecaption{30}{Isoscalar factors for $\bar{\bf 21}$ $\mult$ $\bar{\bf 3}$.}
\begin{center}
%
\end{center}


\spacebetweentables
\tablecaption{31}{Isoscalar factors for $\bar{\bf 21}$ $\mult$ {\bf 6}.}
\begin{center}
%
\end{center}


\spacebetweentables
\tablecaption{32}{Isoscalar factors for $\bar{\bf 21}$ $\mult$ $\bar{\bf 6}$.}
\begin{center}
%
\end{center}


\spacebetweentables
\tablecaption{33}{Isoscalar factors for $\bar{\bf 21}$ $\mult$ {\bf 8}.}
\begin{center}
%
\end{center}

\spacebetweentables
\tablecaption{35}{Isoscalar factors for $\bar{\bf 24}$ $\mult$ {\bf 3}.}
\begin{center}
%
\end{center}


\newpage

\spacebetweentables
\tablecaption{36}{Isoscalar factors for $\bar{\bf 24}$ $\mult$ $\bar{\bf 3}$.}
\begin{center}
%
\end{center}


\newpage

\spacebetweentables
\tablecaption{38}{Isoscalar factors for $\bar{\bf 24}$ $\mult$ {\bf 8}.}
\begin{center}
%
\end{center}


\newpage

\spacebetweentables
\tablecaption{44}{Isoscalar factors for {\bf 35} $\mult$ {\bf 3}.}
\begin{center}
%
\end{center}


\newpage

\spacebetweentables
\tablecaption{45}{Isoscalar factors for {\bf 35} $\mult$ $\bar{\bf 3}$.}
\begin{center}
%
\end{center}


\newpage

\spacebetweentables
\tablecaption{46}{Isoscalar factors for {\bf 35} $\mult$ {\bf 6}.}
\begin{center}
%
\end{center}

\end{document}